\documentclass[traditabstract]{aa}
\usepackage{graphicx}
\usepackage{subcaption}
\usepackage{amsmath}
\usepackage{txfonts}
\usepackage{wasysym}
\usepackage{natbib}
\usepackage[absolute,overlay]{textpos}
\usepackage{tikz}
\usepackage{xcolor}
\usepackage{multirow}

\usetikzlibrary{fit}

\newcommand{\hs}{\hspace*{0.1cm}}
\newcommand{\pymiedap}{\textsc{PyMieDAP} }
\newcommand{\pymiedapx}{\textsc{PyMieDAP}}

\begin{document}


\title{\pymiedap: a Python--Fortran tool to
       compute fluxes and polarization signals of (exo)planets}

\author{Lo\"ic Rossi
		\and Javier Berzosa-Molina
        \and Daphne M. Stam
    }

\offprints{L. Rossi, \email{l.c.g.rossi@tudelft.nl}}

\institute{Faculty of Aerospace Engineering, Technical University Delft,
           Kluyverweg 2, 2629 HS Delft, The Netherlands}

\date{Received xx yy 2018 / Accepted xx yyy 2018}


\abstract{%
\pymiedap (the Python Mie Doubling-Adding Programme) is a Python--based
tool for computing the total, linearly, and circularly polarized fluxes 
of incident unpolarized sun- or starlight that is reflected by, respectively, 
Solar System planets or moons, or exoplanets at a range of wavelengths. 
The radiative transfer computations are based on an adding--doubling
Fortran algorithm and fully include polarization for all orders of scattering.
The model (exo)planets are described by a model atmosphere composed of a stack 
of homogeneous layers containing gas and/or aerosol and/or cloud particles bounded 
below by an isotropically, depolarizing surface (that is optionally black).
The reflected light can be computed spatially--resolved and/or disk--integrated. 
Spatially--resolved signals are mostly representative for 
observations of Solar System planets (or moons), while disk--integrated signals
are mostly representative for exoplanet observations.
\pymiedap is modular and flexible, and allows users to adapt and optimize 
the code according to their needs. 
\pymiedap keeps options open for connections with external
programs and for future additions and extensions.
In this paper, we describe the radiative transfer algorithm that 
\pymiedap is based on and the code's principal functionalities.
And we provide benchmark results of \pymiedap that can be used for testing
its installation and for comparison with other codes. 
\pymiedap is available online under the GNU GPL license at \protect\url{http://gitlab.com/loic.cg.rossi/pymiedap}  
}


\titlerunning{\pymiedap: a tool to compute fluxes and polarization signals of (exo)planets}
\authorrunning{Rossi et al.}

\maketitle



\section{Introduction}
\label{sect1}

Light is usually described only by its total flux, and usually, the total flux
is the only parameter that is measured when observing sunlight that is reflected 
by the Earth or other planets in the Solar System. 
A full description of light, however, requires its state of polarization. 
The state of polarization includes the degree of polarization, $P$, i.e.\ the
ratio of the polarized flux to the total --polarized plus unpolarized-- flux, 
and the direction of polarization, $\chi$. 
The polarized flux can be subdivided into
the linearly polarized flux and the circularly polarized flux. Similarly,
the degree of polarization $P$ can be subdivided into the degree of linear
polarization $P_{\rm l}$ and the degree of circular polarization $P_{\rm c}$.
An excellent description of the polarization of light that is 
reflected by planetary atmospheres and surfaces can be found in
\citet{1974SSRv...16..527H}. That paper also includes 
various methods to compute the state of polarization of light
that is reflected by a planet, amongst others the so--called adding--doubling 
method, that is employed by \pymiedapx, the code that is the topic of this
paper.

There are several reasons for measuring the state of polarization of sunlight
that is reflected by a planet, or starlight that is reflected by an exoplanet.
A first advantage of polarimetry comes from the fact that the light of a
solar type star can be considered to be unpolarized
when integrated across the stellar disk. 
This is known to be true for our Sun \citep[][]{1987Natur.326..270K}, and is
also supported by other studies of nearby FGK stars: for example, \citet{Cotton2017} show
that while active stars can present polarizations up to 45 ppm,
non-active stars have very limited and practically negligeable intrinsic
polarisation.

Meanwhile, light that is scattered within a planetary atmosphere and/or that is reflected 
by the planetary surface will usually be (linearly) polarized.  
For exoplanets, polarimetry could thus allow distinguishing the very faint
planetary signal from the much brighter stellar light. In addition, the
measurement of a polarized signal would immediately confirm the nature
of the object.
\citet{2000ApJ...540..504S} present numerically simulated degrees of (linear)
polarization of the combined light of a star and various types of orbiting
gaseous planets. 
When planets are spatially unresolved from their star, the degree of 
polarization of the system as a whole is the ratio of the polarized planetary 
flux to the sum of the total planetary flux and the stellar flux. 
This degree of polarization is obviously very small, 
i.e.\ on the order of $10^{-6}$ \citep[][]{2000ApJ...540..504S}. 
\citet{2004A&A...428..663S} present numerically simulated degrees of
(linear) polarization of spatially resolved gaseous planets.
For these planets, the degree of polarization can reach several tens of 
percent, depending on the physical properties of the planet and the planetary
phase angle, because they do not include the huge, unpolarized stellar signal. 
Clearly, the degree of polarization that can be observed for a given exoplanet
will depend not only on the intrinsic degree of polarization of the planet, but
also on the stellar flux in the background of the planetary signal. 

Another interesting use of polarimetry is the characterization of the planetary object. 
The degree and direction of polarization of light that has been
scattered by a planet (locally or disk--integrated) is very sensitive to the
properties of the atmospheric scatterers (size, shape, composition),
their spatial distribution, and/or to
the reflection properties of the planetary surface (bidirectional reflection,
albedo) \citep[see e.g.][and references
therein]{1974SSRv...16..527H,2002sael.book.....M}, if present. 
In particular, multiple scattering of light randomizes and hence depolarizes
the light, and adds mainly unpolarized light to the reflected signal. 
The angular dependence of the degree and direction of polarization of the
reflected signal thus preserves the angular patterns of the light that is
singly scattered by the atmospheric particles or the surface, and that is
characteristic for the microphysical properties of the scatterers.  
A famous application of the use of polarimetry for the characterization of a
planetary atmosphere is the retrieval of the composition and size of the cloud
particles in Venus' upper clouds using disk--integrated polarimetry 
(with Earth--based telescopes) at a 
range of phase angles and several wavelengths \citep{Hansen1974}. This
information could not be derived from total flux measurements, because the total
flux is less sensitive to the composition and size of the scattering
particles. Indeed, various types of cloud particles would provide a fit
to the total flux measurements.

Even if one were not interested in measurements of polarization and the
analysis of polarization data,
there are compelling reasons to include polarization in the computation of total
fluxes.

Firstly, because light is fully described by a vector and scattering processes
by matrices, ignoring polarization can induce errors up to several percent 
in computed total fluxes both locally and disk--integrated
\citep{1994JQSRT..51..491M,2005A&A...444..275S}. In particular, in 
gaseous absorption bands, where the linear polarization usually differs
from the polarization in the continuum 
\citep{Fauchez2017, Boesche2009,1994STAMMES,1997ABEN}, such errors will lead to errors
in derived gas mixing ratios and e.g.\ cloud top altitudes 
\citep{2005A&A...444..275S}.

Secondly, many spectrometers are sensitive to the state of polarization 
of the incoming light because of the optical properties of mirrors
and e.g.\ gratings. Knowing the polarization sensitivity of your 
instrument, for example, through calibration in an optical laboratory,
is not sufficient to correct total flux measurements for the state
of polarization of the incoming light if the polarization
of the light is not known \citep{2000JGR...10522379S}. However, one could
include the computed state of polarization of the observed light,
combined with the instrument's polarization sensitivity
in the retrieval of the total fluxes.  

A reason why polarization is usually ignored in radiative transfer 
computations 
is probably that codes that fully include polarization for all orders
of scattering are more complex than codes that ignore polarization,
because the latter treat light as a scalar and scattering processes
as described by scalars, 
while the former have to use vectors and matrices. Polarized radiative
transfer codes therefore usually also require more computation time than scalar
codes. 

\begin{figure*}
\begin{minipage}{\linewidth}
\tikzstyle{fortran} = [draw, fill=blue!10, rectangle, minimum width=3.5cm]
    \tikzstyle{python} = [draw, fill=red!10, rectangle, minimum width=3.5cm]
    \centering
    \begin{tikzpicture}[font=\small,scale=0.9,transform shape]

        \coordinate (Out) at (0,0);
        \coordinate (Dat) at (6,0);
        \coordinate (Pla) at (-6,0);

        \coordinate (Ms) at (0,5);
        \coordinate (Mod) at (-6,5);
        \coordinate (Sur) at (6,5);

        \coordinate (Aer) at (-6,10);
        \coordinate (Ss) at (0,10);

        \coordinate (leg) at (6,9);

        \node[fortran] (output) at (Out) {
            \begin{minipage}[]{0.25\linewidth}
                \footnotesize
                \textbf{Output}

                Stokes I, Q, U and V
                \begin{itemize}
                    \item Resolved on the disk
                    \item Disk integrated
                \end{itemize}
            \end{minipage}
        };
        \node[python] (data) at (Dat) {
            \begin{minipage}[]{0.28\linewidth}
                \footnotesize
                \textbf{Data}

                Geometry of observation
                \begin{itemize}
                    \item SZA, emission, phase, azimuth,\dots
                    \item Latitude, longitude, local time,\dots
                \end{itemize}
            \end{minipage}
        };
        \node[python] (planet) at (Pla) {
            \begin{minipage}[]{0.25\linewidth}
                \footnotesize
                \textbf{Cloud coverage}
                \begin{itemize}
                    \item Homogeneous
                    \item Inhomogeneous
                        \begin{itemize}
                            \item Polar caps
                            \item Bands
                            \item Subsolar cloud
                            \item Patchy clouds
                        \end{itemize}
                \end{itemize}
            \end{minipage}
        };

        \node[python] (aerosols) at (Aer) {
            \begin{minipage}[]{0.25\linewidth}
                \footnotesize
                \textbf{Aerosol properties}
                \begin{itemize}
                    \item $n_r$, $n_i$ (inner, outer)
                    \item $r_{\rm eff}$ (inner, outer), $\nu_{\rm eff}$
                    \item mixing
                \end{itemize}
            \end{minipage}
        };

        \node[fortran] (singsca) at (Ss) {
            \begin{minipage}[]{0.25\linewidth}
                \footnotesize
                \textbf{Single scattering}
                \begin{itemize}
                    \item Mie scattering
                    \item Layered spheres
                \end{itemize}
            \end{minipage}
        };

        \node[fortran] (multsca) at (Ms) {
            \begin{minipage}[]{0.25\linewidth}
                \footnotesize
                \textbf{Multiple scattering}
        
                with doubling-adding method
            \end{minipage}
        };

        \node[python] (surface) at (Sur) {
            \begin{minipage}[]{0.25\linewidth}
                \footnotesize
                \textbf{Surface}
                \begin{itemize}
                    \item Lambertian (non polarizing)
                \end{itemize}
            \end{minipage}
        };

        \node[python] (model) at (Mod) {
            \begin{minipage}[]{0.25\linewidth}
                \footnotesize
                \textbf{Model atmosphere}
                \begin{itemize}
                    \item Planet properties
                        \begin{itemize}
                            \item $g$ 
                        \end{itemize}
                    \item Gas properties
                        \begin{itemize}
                            \item Molecular mass, depolarization, refractive
                                index
                        \end{itemize}
                    \item Layers definition
                        \begin{itemize}
                            \item $\tau$, $P$, $T$
                            \item aerosols
                        \end{itemize}
                \end{itemize}
            \end{minipage}
    };

        \node[draw, minimum width=3cm] (legend) at (leg) {
            \begin{minipage}[]{0.25\linewidth}
                \footnotesize
                \textbf{Legend}
                \begin{itemize}
                    \item \colorbox{red!10}{Python code}
                    \item \colorbox{blue!10}{Fortran code}
                \end{itemize}
            \end{minipage}
        };

    \draw[->, thick] (model) -- (multsca);
    \draw[<-, thick] (output) -- (data);
    \draw[->, thick] (planet) -- (output);
    \draw[->, thick] (surface) -- (multsca);
    \draw[->, thick] (aerosols) -- (singsca);
    \draw[->, thick] (singsca) -- (multsca) node[midway, fill=white] {Expansion
    coefficients};
    \draw[->, thick] (multsca) -- (output) node[midway, fill=white] {Fourier coefficients};
    \end{tikzpicture}
\end{minipage}
\caption{The modules comprising \pymiedapx. 
         The blue boxes represent Fortran codes,
         the red boxes Python codes.}
\label{pymiedap-diagram}
\end{figure*}
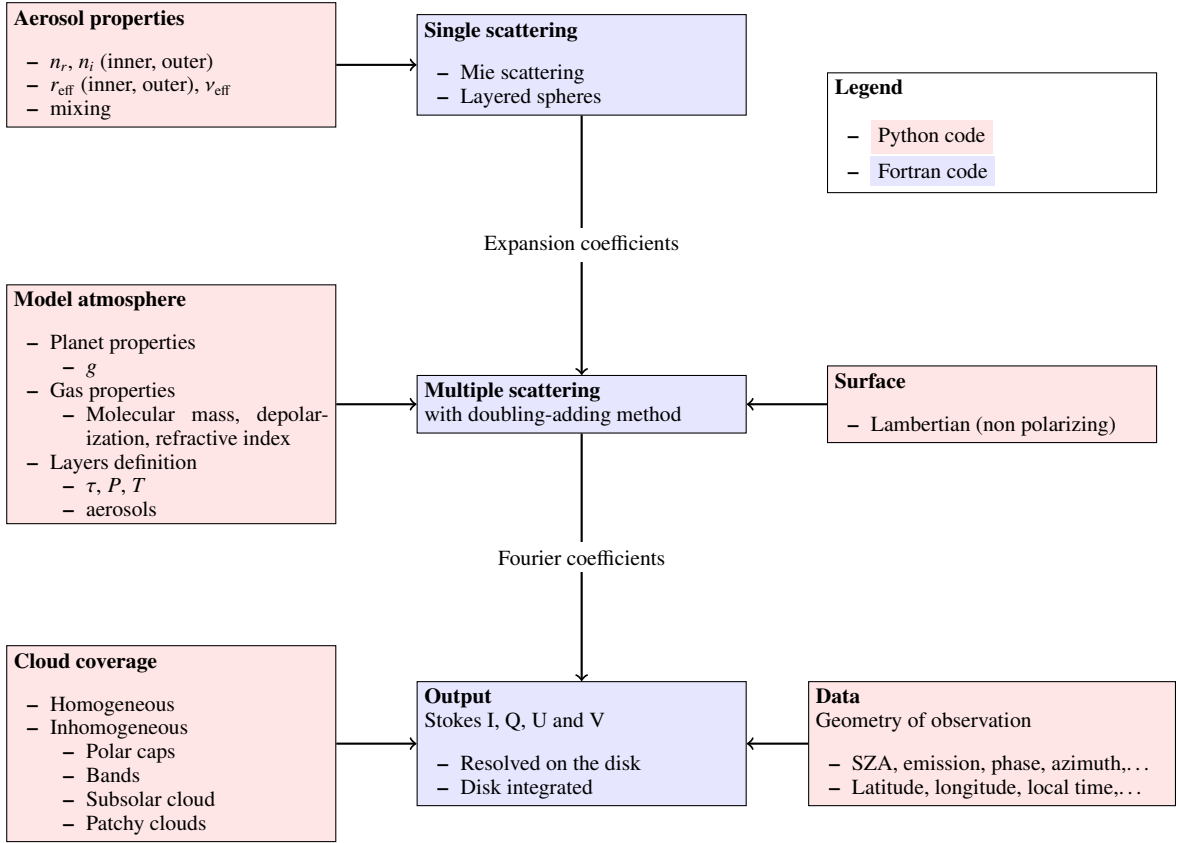

In this paper, we present \pymiedapx, a user--friendly, modular, Python--based 
tool for computing the total and polarized fluxes of light that is reflected by
(exo)planets.\footnote{\pymiedap is available at \url{http://gitlab.com/loic.cg.rossi/pymiedap}.}
The radiative transfer computations in \pymiedap are performed 
with an adding--doubling algorithm written in Fortran, 
as described by \citep{1987A&A...183..371D}, while input and output are
handled with Python code.
Figure~\ref{pymiedap-diagram} provides a view of the modules composing
\pymiedapx. The blue boxes represent codes written in Fortran, and the red
boxes Python code. Arrows indicate interfaces 
using \verb,f2py, \citep{Peterson2009}.
Each \pymiedap module will be described in more detail in this paper.

The structure of this paper is as follows.
Section~\ref{definitions-flux-pol} provides the 
definitions of the vector elements that describe the state of polarization 
of light as used in \pymiedapx.
Section~\ref{reflected-light} contains the formulae required to calculate the
Stokes vectors of sun- or starlight that is locally reflected by a region on a planet
for a range of illumination and viewing geometries.
The components of the Fourier--series decomposition of these vectors are stored 
in files in a database that is accessed to 
compute reflected light vectors for specific geometries.
Section~\ref{single-scat} describes the computation of the single scattering
properties of atmospheric gases and aerosols, and the reflection by the surface.
Section~\ref{dap} describes the adding--doubling radiative transfer algorithm 
used in \pymiedap.
Section~\ref{geometry-masks} presents the method used to compute the geometries
for locally reflected light, and describes how previously stored database 
files are used to compute the locally reflected light vector as well as 
how to integrate these locally reflected light vectors across the 
illuminated and visible part of a planetary disk, in
order to obtain the disk--integrated Stokes vector. 
In Sect.~\ref{benchmark}, we compare reflected Stokes vectors obtained 
with \pymiedap against previously published results obtained using a 
similar adding--doubling radiative transfer algorithm without the intermediate
step of using pre--calculated database files, and without the Python shell.
Section~\ref{sect-summary} summarizes the paper and discusses future work.
Appendix~\ref{fourier-files} provides a detailed description of the 
format of the database files that is used in the codes.
Appendix~\ref{appendix-angles} provides equations for the computation of some angles used in the code.


\section{Defining the flux and polarization of light}
\label{definitions-flux-pol}

The radiance ('intensity') and state of polarization of a quasi--monochromatic
beam of light can be described by a Stokes vector ${\bf I}$ as follows 
\citep[see, e.g.][]{1974SSRv...16..527H,2004Hovenier}
\begin{equation}
   {\bf I}= \left[ \begin{array}{c}
          I \\ Q \\ U \\ V \end{array} \right].
\label{eq_fluxvector1}
\end{equation}
Here, Stokes parameter $I$ is the total radiance, $Q$ and $U$ describe the linearly
polarized radiances, and $V$ the circularly polarized radiance.  
All four parameters have the dimension W m$^{-2}$ sr$^{-1}$ (or W m$^{-3}$
sr$^{-1}$ if taken as functions of the wavelength $\lambda$).  
We will also use the irradiance or flux vector 
$\pi {\bf F}= \pi [F, Q, U, V]$, of which all parameters have the 
dimension W m$^{-2}$ (or W m$^{-3}$ if taken as functions of $\lambda$). 
Unless specified otherwise, the equations in this paper also hold for 
flux vector $\pi {\bf F}$.

Parameters $Q$ and $U$ are defined with respect to a reference plane.  
We use two types of reference planes:
\begin{enumerate}
\item {\em The local meridian plane}, which contains the local zenith direction and the 
direction of propagation of the light. The local meridian plane is used in the
computation of $Q$ and $U$ of locally reflected light.
\item {\em The planetary scattering plane}, which contains the centre 
of the planet, and the directions to the centre of the star and to the observer.
This plane is mainly used to define 
$Q$ and $U$ of light that has been reflected by the planet as a whole, e.g.\
for simulating signals of (spatially unresolved) exoplanets. 
\end{enumerate}
Parameters $Q$ and $U$ can be
transformed from one reference plane to another 
using a so--called rotation matrix ${\bf L}$ \citep[see][]{1983A&A...128....1H}
\begin{equation}
   {\bf L}(\beta)= \left[ \begin{array}{cccc}
             1 & 0 & 0 & 0 \\
             0 & \cos 2\beta & \sin 2\beta & 0 \\
             0 & -\sin 2\beta & \cos 2\beta & 0 \\
             0 & 0 & 0 & 1 \\
              \end{array}
              \right],
\label{eq_L}
\end{equation}
with $\beta$ the angle between the two reference planes, measured rotating in
the anti--clockwise direction from the old to the new reference plane when
looking towards the observer ($\beta \geq 0^\circ$).

\begin{figure}[h]
\centering
\includegraphics[width=0.8\linewidth]{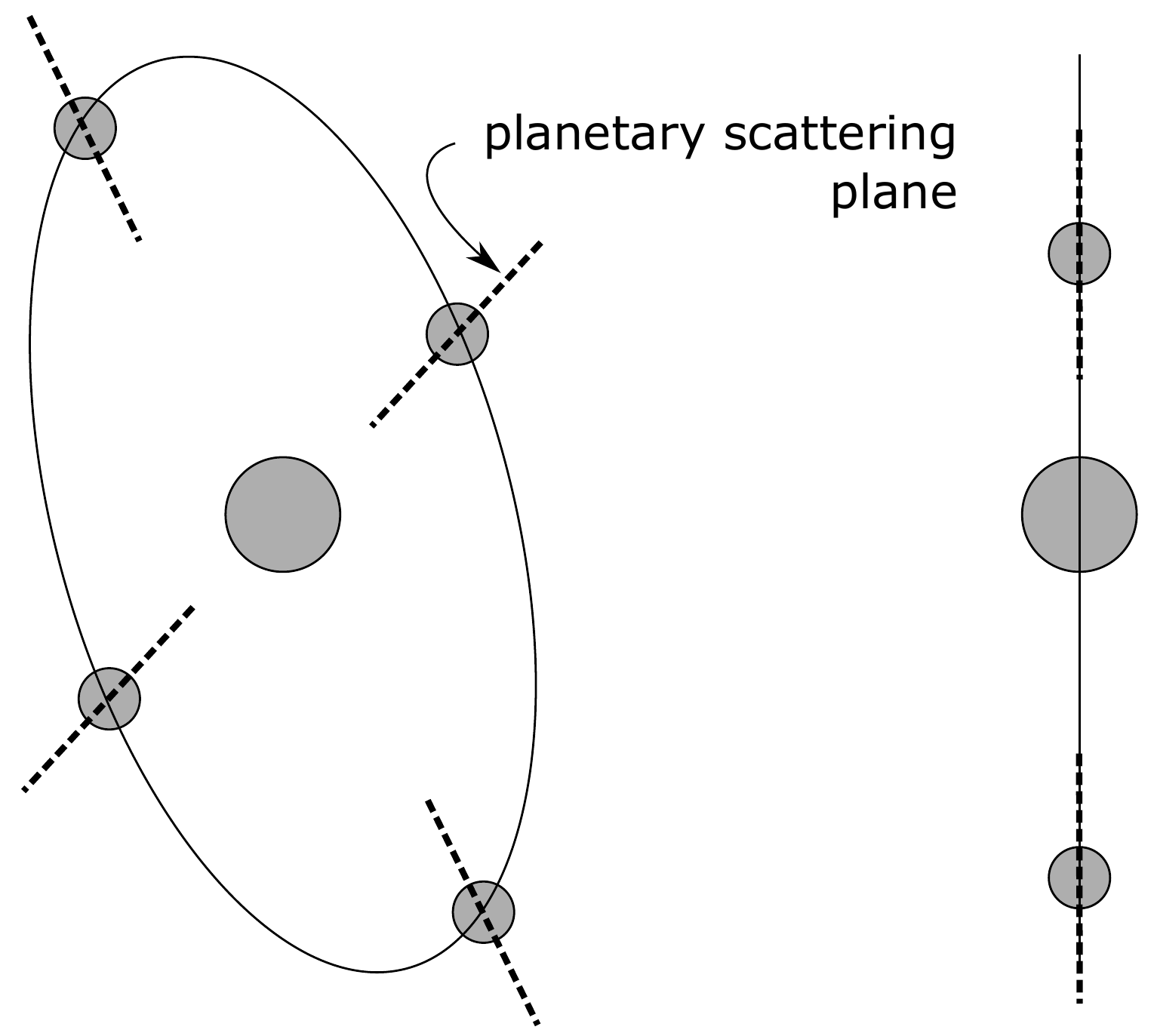}
\caption{Illustration of the rotation of the planetary scattering plane for an orbit with a random orientation (left) and with edge-on configuration (right). The first case would require further rotations to express the Stokes elements in the reference plane of the observer.}
\label{fig:rot-scat-plane}
\end{figure}

In \pymiedapx, the default reference plane for local reflections and 
disk--integrated reflected light, is the planetary scattering plane. 
For locally reflected light, the vector that is computed with respect to the
local meridian plane is rotated to be defined with respect to the planetary 
scattering plane before being provided as output.
As a planet orbits its star, the planetary scattering plane will usually 
rotate on the sky as seen from the observer, except if the orientation
of the orbit is edge--on with respect to the observer (see Fig. \ref{fig:rot-scat-plane}).
By applying additional rotations, Stokes vectors defined with respect 
to the planetary scattering plane can straightforwardly be redefined  
to e.g.\ the optical plane of an instrument or the detector 
\citep[for a detailed description of these rotations, see][]{Rossi2017}.

The degree of polarization of the beam of light described by vector ${\bf I}$
(Eq.~\ref{eq_fluxvector1}) is defined as
\begin{equation}
    P = \frac{\sqrt{Q^2 + U^2 + V^2}}{I}.
\label{eq_polt}
\end{equation}
The degree of {\em linear} polarization is defined as 
\begin{equation}
    P_{\rm l} = \frac{\sqrt{Q^2 + U^2}}{I},
\label{eq_pol}
\end{equation}
and the degree of {\em circular} polarization as
\begin{equation}
    P_{\rm c} = \frac{V}{I}.
\label{eq_polc}
\end{equation}

While the degree of linear polarization is independent of the choice of 
reference plane for Stokes parameters $Q$ and $U$, the direction or angle of linear 
polarization, $\chi$, is not independent of the choice of reference plane.
Angle $\chi$ can be derived from
\begin{equation}
   \tan 2\chi = U/Q.
\end{equation}
The value of $\chi$ is chosen in the interval $[0^\circ,180^\circ \rangle$, and such 
that $\cos 2\chi$ has the same sign as $Q$ \citep[see][]{1974SSRv...16..527H}. 

If $U=0$, the direction of polarization of the light is either perpendicular
($Q < 0$, $\chi=90^\circ$) or parallel ($Q > 0$, $\chi=0^\circ$) to the 
reference plane. 
In that case, we can use an alternative definition of the degree of linear
polarization that captures the information about the direction of polarization,
namely
\begin{equation}
   P_{\rm ls} = -\frac{Q}{I},
\label{eq_sign}
\end{equation}
where $P_{\rm ls} > 0$ ($P_{\rm ls} < 0$) corresponds with light 
that is polarized perpendicularly (parallel) to the reference plane.

Regarding the circular polarization (Eq.~\ref{eq_polc}), 
our convention for the sign is as follows:
$V$ and thus $P_{\rm c}$ is positive when the observer ’sees’
the electric vector of the light rotating in the anti–clockwise direction, 
and $V$ and thus $P_{\rm c}$ is negative, when 
the observer ’sees’ the vector rotating in the clockwise direction.


\section{Calculating reflected light}
\label{reflected-light}

With \pymiedapx, one can calculate the Stokes vector ${\bf I}$ 
(cf.\ Eq.~\ref{eq_fluxvector1}) of light that is locally reflected by a planet.
Here, we refer to {\em locally} reflected light if a single combination of  
illumination and viewing geometries involved in the reflection process applies.
With \pymiedapx, one can also integrate Stokes vectors of locally reflected light 
across the planet, taking into account variations of the
atmospheric properties and/or surface albedo, as well as the variations 
of the illumination and viewing geometries.

Below, we explain the calculation of the locally reflected light (Sect.~\ref{sect3.1})
and the integration of locally reflected light 
across the illuminated and visible part of a planetary disk (Sect.~\ref{sect3.2}).
The integration over a smaller part of a planet could straightforwardly
be derived from the latter explanation\footnote{This has not yet been implemented 
in PyMieDAP}.


\subsection{Calculating locally reflected light}
\label{sect3.1}

We calculate a locally reflected vector ${\bf I}$ (see Eq.~\ref{eq_fluxvector1}) 
according to \citep[see][]{1974SSRv...16..527H}
\begin{equation}
   {\bf I}(\mu,\mu_0,\phi-\phi_0,\lambda)= 
           \mu_0 \hs {\bf R}(\mu,\mu_0,\phi-\phi_0,\lambda) \hs {\bf F}_0(\lambda),
\label{eq_reflvector}
\end{equation}
with ${\bf F}_0$ the vector describing the incident light and
${\bf R}$ the $4 \times 4$ local planetary reflection matrix.
The reference plane for ${\bf I}$ is the local meridian plane, the plane
containing the local zenith direction and the propagation direction of 
the reflected light
(see Sect.~\ref{definitions-flux-pol}).

Furthermore in Eq.~\ref{eq_reflvector}, $\mu = \cos \theta$, with $\theta$ 
the angle between the direction of propagation of the reflected light and the 
upward vertical ($0^\circ \leq \theta < 90^\circ$), and
$\mu_0 = \cos \theta_0$, with $\theta_0$ the angle between 
the upward vertical and the direction to the sun or star
($0^\circ \leq \theta_0 < 90^\circ$). 
The azimuthal difference angle $\phi-\phi_0$ is measured between the two 
vertical planes 
containing the directions of propagation of the reflected and the incident light, 
respectively ($0^\circ \leq \phi-\phi_0 \leq 180^\circ$). 
To get our definition for $\phi-\phi_0$ clear, consider
light that is reflected in the vertical plane that contains the 
local zenith direction, 
and the direction towards the sun or star. If the reflected light propagates 
in the half of the vertical plane that contains the sun or star, 
$\phi-\phi_0=180^\circ$. If the reflected light propagates in the other 
half of the plane, $\phi-\phi_0=0^\circ$. 

In \pymiedapx, it is assumed that the incident sun or starlight is unpolarized
\citep{1987Natur.326..270K}, although this assumption is not inherent to the radiative
transfer algorithm \citep{1987A&A...183..371D}, and \pymiedap could 
easily be adapted for polarized incident light.
Vector ${\bf F}_0$ of the light that is incident on a model planet thus equals the column 
vector $[F_0, 0, 0, 0]$ or $F_0 [1, 0, 0, 0]$, where $F_0$ equals 
the total incident solar/stellar flux measured perpendicular to the direction 
of incidence {\em divided by $\pi$} \citep[see][]{1974SSRv...16..527H}.
For example, if the total incident flux measured perpendicular
to the direction of incidence equals $S_0$ W~m$^{-2}$, 
$F_0= S_0/\pi$ W~m$^{-2}$.

With the assumption of unpolarized incident sun or starlight, 
only the elements of ${\bf R}_1$, the first column of the $4 \times 4$ 
local planetary reflection matrix ${\bf R}$
are relevant for the calculation of the locally reflected vector 
${\bf I}$, since Eq.~\ref{eq_reflvector} then transforms into
\begin{equation}
   {\bf I}(\mu,\mu_0,\phi-\phi_0,\lambda)= 
           \mu_0 \hs {\bf R}_1(\mu,\mu_0,\phi-\phi_0,\lambda) \hs F_0(\lambda).
\label{eq_reflvector2}
\end{equation}
The local reflection vector ${\bf R}_1$ depends on the illumination and
viewing geometries and the properties of the local planetary 
atmosphere and surface. 
The user can provide \pymiedap with a list of illumination and viewing 
geometries, e.g. geometries that pertain to observations from a satellite
that orbits a planet.
Given the properties of the local atmosphere and surface, the calculation
of ${\bf R}_1$ and subsequently ${\bf I}$
is performed by \pymiedap as described in Sect.~\ref{dap}.
Note that locally reflected light vector ${\bf I}$ as described by Eq.~\ref{eq_reflvector2}
is defined with respect to the local meridian plane. \pymiedap will redefine it
with respect to the planetary scattering plane by calculating the local 
angle $\beta$ and applying the rotation matrix ${\bf L}$ as 
defined in Eq.~\ref{eq_L}.

In case circular polarization is ignored, vector ${\bf R}_1$
and reflected light vector ${\bf I}$ each
comprise only 3~elements. In case circular and linear polarization are both
ignored, ${\bf R}_1$ and ${\bf I}$ are scalars, and Eq.~\ref{eq_reflvector2}
could be written as
\begin{equation}
    I(\mu,\mu_0,\phi-\phi_0,\lambda)= 
      \mu_0 \hs R_1(\mu,\mu_0,\phi-\phi_0,\lambda) \hs F_0(\lambda).
\label{eq_reflvector3}
\end{equation}
Contrary to ignoring linear polarization,
ignoring circular polarization usually only leads to very small errors in the
computed total and linearly polarized fluxes \citep{2005HovenierStam}.

In the following, we will assume polarization (both linear and circular) is 
taken into account and use vectors and matrices instead of scalars.


\subsection{Calculating disk--integrated reflected light}
\label{sect3.2}

To calculate signals of spatially unresolved planets, such as exoplanets,
we integrate the locally reflected starlight as given by Eqs.~\ref{eq_reflvector2}
and~\ref{eq_reflvector3}, 
over the illuminated and visible part of the planetary disk,
according to \citep[see][Eq.~16]{Stam2006}
\begin{eqnarray}
  \pi {\bf F}(\alpha,\lambda) & = & 
    \frac{1}{d^2} \int_{\rightmoon} \mu {\bf L}(\beta) 
                  {\bf I}(\mu,\mu_0,\phi-\phi_0,\lambda) 
                  dO \label{eq_fl1} \\
            & = & \frac{1}{d^2} \int_{\rightmoon} \mu \mu_0 
                   {\bf L}(\beta) {\bf R}_1(\mu,\mu_0,\phi-\phi_0,\lambda) 
                   F_0(\lambda) dO. 
\label{eq_fluxtot}
\end{eqnarray}
Here, $\pi {\bf F}$ is the flux vector of the reflected starlight as it arrives at the 
observer located at a distance $d$ from the planet, with $\pi F$ the flux
measured perpendicularly to the direction of propagation of the light.
Furthermore, $\mu \hspace*{0.1cm} dO/d^2$ is the solid angle under which surface area 
$dO$ on the planet is seen by the observer ($\mu = \cos \theta$). 
The planet's radius $r$ is incorporated in the surface integral
(the planet is thus not assumed to be a unit sphere).
Reflected light vector $\pi {\bf F}$ depends on the planetary phase 
angle $\alpha$, 
i.e. the angle between the star and the observer as measured from the centre of 
the planet ($0^\circ \leq \alpha \leq 180^\circ$). The range of observable phase angles for an exoplanet will depend on the orbital inclination and/or on the inner working angle of the instrument.

Furthermore in Eq.~\ref{eq_fluxtot},
each locally reflected vector ${\bf I}$, and hence each local reflection matrix
${\bf R}_1$ is rotated such that the reference
plane is no longer the local meridian plane, but the planetary scattering plane.
The local rotation angle $\beta$ depends on the local viewing angle $\theta$ 
and the location of surface area $dO$ on the planet.

The geometric albedo $A_{\rm G}$ of the planet with radius $r$ at distance $d$
is given by
\begin{equation}
   A_{\rm G}(\lambda)= \frac{\pi F(0^\circ,\lambda)}{\pi F_0}(\lambda) \frac{d^2}{r^2},
\label{eq_geoalb}
\end{equation}
where $\pi F_0(0^\circ,\lambda)$ is the reflected total flux at wavelength 
$\lambda$ and phase angle $\alpha$ equal to $0^\circ$.

In \pymiedapx, the integration in Eq.~\ref{eq_fluxtot} is replaced by a 
summation over locally reflected Stokes vectors. In order to do so, we
divide the planetary disk on the sky in pixels,
and compute the locally reflected Stokes vector at the centre of 
each pixel.
A pixel contributes to the disk--signal when its center is within the disk--radius.
The integration is then given by:
\begin{equation}
   \pi {\bf F}(\alpha,\lambda) = \frac{F_0(\lambda)}{d^2} 
       \sum_{n=1}^{N} \mu_n \hs \mu_{0n} \hs {\bf L}(\beta_n) \hs 
         {\bf R}_1(\mu_n,\mu_{0n},\phi_n-\phi_{0n},\lambda) \hs dO_n, 
\label{eq_sum}
\end{equation}
with $N$ the number of illuminated and visible pixels on the planetary disk,
and with subscript $n$ indicating that
$\mu$, $\mu_0$, $\phi-\phi_0$, and $\beta$ 
depend on the location of the pixel on the planet.
In addition, $\mu_0$, $\phi-\phi_0$, and $\beta$ at a given location
of the pixel on the planet depend on the planet's phase angle $\alpha$ (Appendix \ref{appendix-angles} provides relations that can be used to derive these angles).  
In the summation, $dO_n$ refers to the area of the pixel as measured on
the surface of the planet.

Although not explicitly indicated in Eqs.~\ref{eq_sum} and~\ref{eq_fluxtot}, 
${\bf R}_1$ will usually also depend on the location (of a pixel) on the planet.
Typical horizontally inhomogeneities would be: the surface coverage and altitude,
and the atmospheric composition and structure.
The obvious horizontal variations on Earth are of course the oceans and
the continents, and in the atmosphere, the clouds.
Horizontal inhomogeneities can be taken into account by using different
local reflection vectors ${\bf R}_1$ across the planet (in that case,
${\bf R}_1$ in Eq.~\ref{eq_sum} would include a subscript $n$).

Given a model planet and a planetary phase angle $\alpha$, the steps to 
evaluate Eq.~\ref{eq_sum}, are the following: \begin{enumerate}
\item Divide the planet in pixels small enough to 
be able to assume that the planet properties across each pixel 
are horizontally homogeneous, and to be able to follow the limb and the
terminator of the planet.
\item Calculate for (the centre of) each pixel the angles $\mu$ (i.e. $\cos \theta$), 
$\phi$, and the rotational angle $\beta$. These angles are independent of the 
location of the sun or star with respect to the planet. 
\item Calculate for the given phase angle $\alpha$, 
and for (the centre of) each pixel, $\mu_0$ (i.e. $\cos \theta_0$) and $\phi_0$.
These angles depend on the location of the sun or star with respect 
to the planet. 
\item Calculate for (the centre of) each pixel, the 
column vector ${\bf R}_1$ of the locally reflected light 
using the appropriate database file (see Sect.~\ref{dap}).
\item Perform the summation described by Eq.~\ref{eq_sum}.
\end{enumerate}

\begin{figure}
\centering
\includegraphics[width=8.0cm]{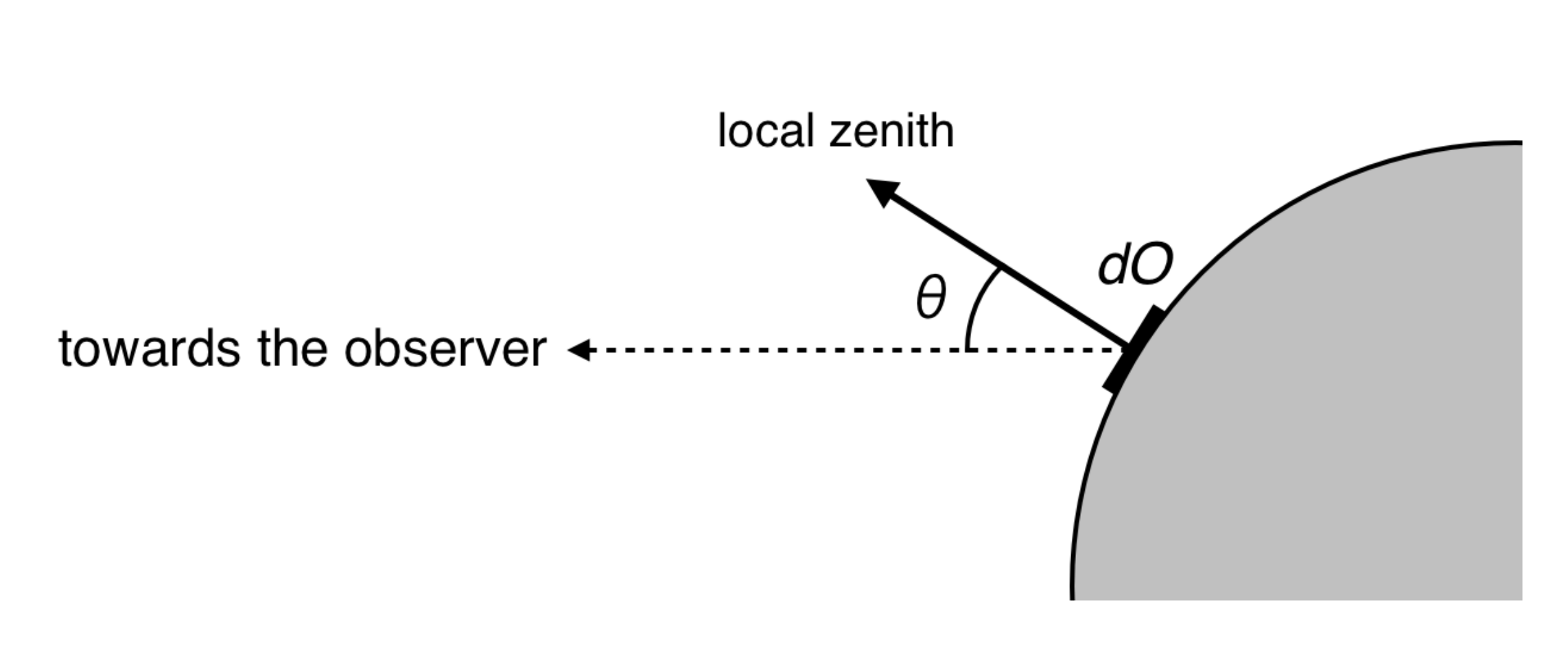}
\caption{Sketch of a surface area $dO$ on the planet (side view) and its 
         projection towards the observer. The observer 'sees' a pixel area
         equal to $\mu \hs dO = \cos \theta \hs dO$, with $\theta$ the local
         viewing zenith angle.}
\label{fig_dO}
\end{figure}

The pixels can be defined on
the planet, for example by using a latitude and longitude grid, in which case 
$\mu \hs dO$, the projected area of the pixel (see Eq.~\ref{eq_fluxtot}), 
i.e.\ the pixel area as 'seen' by the observer (see Fig.~\ref{fig_dO})
will depend on the location of the pixel on the planet. 
\pymiedap uses a grid of equally sized 
square pixels, similar to detector pixels, and uses the projections of those pixels
onto the planet to divide the planet into separate regions.
In this case, $\mu \hs dO$ is simply the surface area of the square pixel, and there is
no need to calculate $dO$, the surface of the projected pixel on the planet (which can
have a complicated shape). The result of the integration will depend
on the pixel size, and thus on the number of pixels across the planetary disk,
in particular at large phase angles, where the pixels
should be sufficiently small resolve the crescent shape of the illuminated part of the 
planetary disk. 

\begin{figure}[h]
\centering
\includegraphics[width=\linewidth]{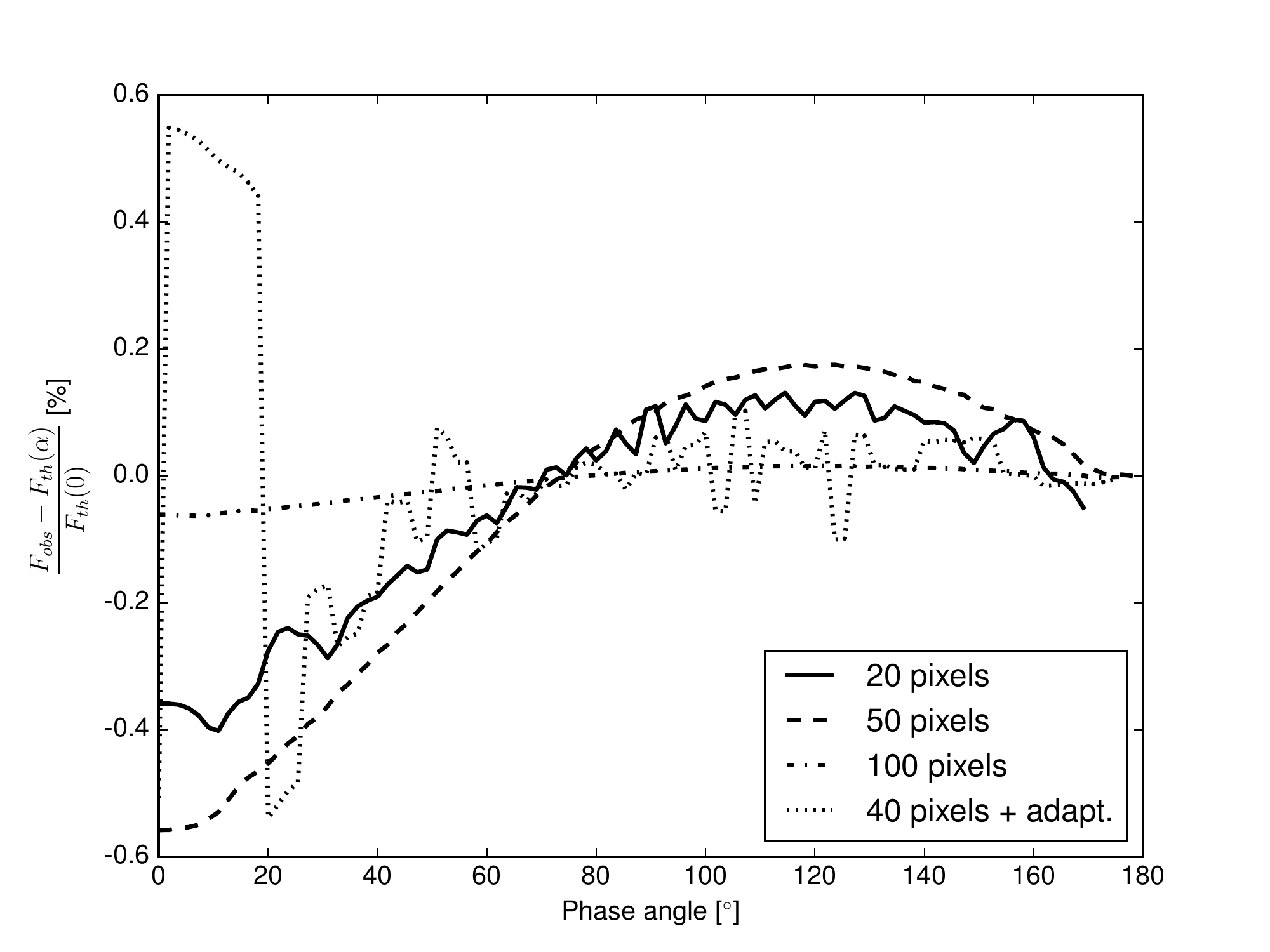}
\caption{The difference between the disk--integrated flux $F$ computed using
         Eq.~\ref{eq_sum} and computed using the analytical expression 
         for a sphere with a Lambertian reflecting surface and a geometric 
         albedo $A_{\rm G}$ of 1.0 \citep[see][and Eq.~\ref{eq_lam}]{Stam2006}, 
         as a function of the
         planet's phase angle $\alpha$ for various values of $N_{\rm eq}$,
         the number of pixels along the planet's equator.
         The difference has been normalized to $A_{\rm G}$.
         The dotted line: $N_{\rm eq}$ increases with $\alpha$ according to
         $N_{\rm eq}(\alpha) = N_{\rm eq}(0^\circ) [ 1 + \sin^2 (\alpha/2)]$,
         with $N_{\rm eq}(0^\circ)= 40$.
         }
\label{fig_surface}
\end{figure}

The computation time increases linearly with $N$, the number of pixels
on the illuminated and visible part of the planetary disk. In order to keep computing
times low, it is thus important to find a balance between the number of pixels and 
the accuracy.  
The relative error in the total flux of a Lambertian reflecting planet
due to the pixel size is shown in Fig.~\ref{fig_surface}.
The errors decrease with increasing value of $N$ at any given phase angle $\alpha$.
For a given value of $N$, the relative error increases with increasing phase angle 
$\alpha$, thus with decreasing width of the planetary crescent, but the total disk--integrated flux also decreases with increasing 
$\alpha$, with $\pi F(180^\circ)= 0.0$. Thus while the relative errors at 
large phase angles can be very large, the absolute errors remain small. 
For computations across a range of phase angles, \pymiedap can automatically increase
the number of pixels across the equator $N_{\rm eq}$ (and therefore $N$) with increasing 
$\alpha$ in order to keep the errors small. The results of
this automatic increase of $N$ are also shown in Fig.~\ref{fig_surface}.
This 'adaptive pixels scheme' can be useful as a trade-off between computational 
efficiency and the need to resolve the planet at large phase angles, as using a 
smaller number of pixels for a full planet is usually acceptable, while it might 
be detrimental to the computed output for thin crescents.

Note that for horizontally homogeneous planets, the Stokes vector of the 
hemisphere above the planetary scattering plane equals that of the southern
hemisphere, except for the sign of Stokes parameters $U$ and $V$. 

For horizontally homogeneous planets, \citet{Stam2006} describe an efficient 
algorithm that does not require dividing the planet into pixels, and that evaluates 
the disk--integrated Stokes vector $\pi {\bf F}$ at arbitrary phase angles $\alpha$.
With this algorithm (that has not been implemented in \pymiedapx), 
vectors of horizontally inhomogeneous planets can be 
approximated using a weighted sum of vectors of horizontally homogeneous 
planets \citep[see][]{2008A&A...482..989S}, 
with the weights depending on the fractions the various inhomogeneities
cover on the illuminated and visible part of the planetary disk. 
With such a weighted sum approximation,
one can estimate the range of signals to be expected from an exoplanet. 
However, because a weighted sum does not account for the actual spatial 
distribution of the inhomogeneities, and e.g.\ the change therein when a 
planet rotates about its axis, it cannot be used for interpreting
signals of planets that are known to exhibit significant horizontal 
inhomogeneities. For such applications, \pymiedapx's pixel approach
should be used.


\section{Describing the model atmosphere and surface}
\label{single-scat}

\pymiedap's adding--doubling radiative transfer algorithm assumes a flat model
atmosphere that is horizontally homogeneous, but that can be vertically inhomogeneous
because different horizontally homogeneous atmospheric layers can be stacked. 
A model atmosphere is bounded below by a flat, horizontally homogeneous
surface. 
Below we describe how the scattering by the gaseous molecules, the aerosol
particles and the reflection by the surface is implemented in \pymiedapx.

\subsection{The model atmosphere}

A model atmosphere consists of a stack of horizontally homogeneous layers.
Each atmospheric layer can contain gaseous molecules and/or aerosol particles
(including cloud particles).
For every layer, the algorithm requires the total 
optical thickness $b$, the single scattering matrix ${\bf F}$ of the gas and/or
aerosol particles in the layer, and their single scattering albedo $a$. 

The single scattering of incident light by gas molecules is described by
anisotropic Rayleigh scattering \citep[][]{1974SSRv...16..527H}, which includes
both the inelastic Cabannes scattering and the elastic Raman scattering
processes \citep[][]{1981ApOpt..20..533Y}.
Although overall energy is thus conserved, narrow spectral features that are
due to Raman scattering, such as the filling--in of absorption lines in stellar
spectra upon inelastic scattering in the planetary atmosphere
\citep[][]{1962Natur.193..762W,2002JGRD.107t.AAC1S} cannot be reproduced with
\pymiedapx's radiative transfer code.

We use the single scattering matrix for anisotropic Rayleigh scattering as 
described by \citet{1974SSRv...16..527H}:
\begin{equation}
   {\bf F}^{\rm m}(\Theta,\lambda) =
          \left[ \begin{array}{cccc}
     A^{\rm m}_1(\Theta,\lambda) & B^{\rm m}_1(\Theta,\lambda) & 0 & 0 \\
     B^{\rm m}_1(\Theta,\lambda) & A^{\rm m}_2(\Theta,\lambda) & 0 & 0 \\
     0 & 0 & A^{\rm m}_3(\Theta,\lambda) & 0 \\
     0 & 0 & 0 & A^{\rm m}_4(\Theta,\lambda)
          \end{array} \right],
\label{eq_scatmat}
\end{equation}
with the superscript '{\rm m}' referring to molecules. 
The single scattering angle $\Theta$ is measured with respect to the direction
of propagation of the incoming beam of light: $\Theta=0^\circ$ for forward and
$180^\circ$ for backward scattered light.

Single scattering matrix ${\bf F}^{\rm m}$ is normalized at every wavelength 
$\lambda$ such that element $A^{\rm m}_1$ (the 'phase function'), averaged
over all scattering directions equals one \citep[see][]{1974SSRv...16..527H}.
The elements of ${\bf F}^{\rm m}$ are the following
[Hansen and Travis, 1974]
\begin{equation}
   A^{\rm m}_1(\Theta,\lambda)= 1 - \frac{1}{4} \Delta(\lambda) (1-3\cos^2 \Theta) 
\label{eq_scatR1}
\end{equation}
\begin{equation}
   A^{\rm m}_2(\Theta,\lambda)= \frac{3}{4} \Delta(\lambda) (1+\cos^2 \Theta) 
\end{equation}
\begin{equation}
   A^{\rm m}_3(\Theta,\lambda)= \frac{3}{2} \Delta(\lambda) \cos \Theta 
\end{equation}
\begin{equation}
   A^{\rm m}_4(\Theta,\lambda)= \frac{3}{2} \Delta(\lambda) \Delta'(\lambda) \cos \Theta
\end{equation}
\begin{equation}
   B^{\rm m}_1(\Theta,\lambda)= - \frac{3}{4} \Delta(\lambda) \sin^2 \Theta,
\label{eq_scatR2}
\end{equation}
with
\begin{equation}
   \Delta(\lambda) = \frac{1 - \rho(\lambda)}{1 + \rho(\lambda)/2} {\hspace{0.5cm} \mbox{and} \hspace{0.5cm}}
   \Delta'(\lambda) = \frac{1 - 2\rho(\lambda)}{1 + \rho(\lambda)/2}.
\end{equation}
While the depolarization factor $\rho$ depends on $\lambda$
\citep[see, e.g.][]{2005JQSRT..92..293S,1984P&SS...32..785B}, for 
most gases, the precise spectral dependence is not well--known.
Typical values for the $\rho$ are 0.09 for CO$_2$ (fairly wavelength
independent) and 0.0213 for N$_2$ (at 500~nm).
The current version of \pymiedap assumes a wavelength independent value for 
$\rho$.

Our adding--doubling radiative transfer algorithm (see Sect.~\ref{dap}), 
does not actually use
the scattering matrix elements themselves, but rather the coefficients of their
expansion in generalized spherical functions, as described in detail by
\citet{1984A&A...131..237D}.
The expansion coefficients for anisotropic Rayleigh scattering are given in
e.g.\ \citet{2002JGRD.107t.AAC1S} and are, for a given value of
$\rho$, computed by \pymiedapx.

With \pymiedapx, the user can directly define $b^{\rm m}$, the 
gaseous extinction optical thickness of an atmospheric layer (measured along the
vertical direction), or the user can specify the pressure difference across 
an atmospheric layer and leave \pymiedap to compute $b^{\rm m}$ for each
given wavelength $\lambda$ under
the assumption of hydrostatic equilibrium, according to 
\begin{equation}
   b^{\rm m}(\lambda)= \sigma^{\rm m}(\lambda) \hspace*{0.1cm} N^{\rm m}
                     = \sigma^{\rm m}(\lambda) \hspace*{0.1cm} N_{\rm A}
                       \frac{p_{\rm bot}-p_{\rm top}}{m g},
\label{eq_bm}   
\end{equation}
with $\sigma^{\rm m}$ the molecular extinction cross--section 
(in $\mu$m$^{2}$ molecule$^{-1}$),
$N^{\rm m}$ the column number density of the gas (in molecules $\mu$m$^{-2}$),
$p_{\rm bot}-p_{\rm top}$ the pressure difference (in bars or 10$^{-5}$ N m$^{-2}$), 
$N_{\rm A}$ the constant of Avogadro (i.e.\ $6.022140857~\cdot~10^{23}$),
$m$ the mass per mole (in atomic mass units or g mole$^{-1}$), 
and $g$ the acceleration of gravity (in m s$^{-2}$). Apart from the pressure
levels, the user specifies both $m$ and $g$. Note that we have left out 
factors to account for unit conversions in Eq.~\ref{eq_bm} and in
equations below.

The molecular extinction cross--section is the sum of the molecular 
scattering and absorption cross--sections, as follows
\begin{equation}
   \sigma^{\rm m}(\lambda) = \sigma^{\rm m}_{\rm sca}(\lambda) +
                             \sigma^{\rm m}_{\rm abs}(\lambda).
\end{equation}
Combining this with Eq.~\ref{eq_bm}, it is clear that
\begin{equation}
   b^{\rm m}(\lambda) = \sigma^{\rm m}_{\rm sca}(\lambda) \hspace*{0.1cm} N^{\rm m} +
                        \sigma^{\rm m}_{\rm abs}(\lambda) \hspace*{0.1cm} N^{\rm m}
                      = b^{\rm m}_{\rm sca}(\lambda) + b^{\rm m}_{\rm abs}(\lambda),
\label{eq_sumbm}                      
\end{equation}
with $b^{\rm m}_{\rm sca}$ and $b^{\rm m}_{\rm abs}$ the layer's molecular scattering 
and absorption optical thicknesses, respectively. 
To include gaseous absorption, the user defines the gaseous absorption optical
thickness $b^{\rm m}_{\rm abs}$ per wavelength.

\pymiedap computes the molecular scattering cross--section $\sigma^{\rm m}_{\rm sca}$
according to
\begin{equation}
    \sigma^{\rm m}_{\rm sca}(\lambda) = \frac{24 \pi^3}{N_{\rm L}^2} 
              \frac{(n^2(\lambda)-1)^2}{(n^2(\lambda)+2)^2}
              \frac{(6 + 3\rho(\lambda))}{(6 - 7 \rho(\lambda))} 
              \frac{1}{\lambda^4},
\label{eq_sigma}              
\end{equation}
with $N_{\rm L}$ Loschmidt's number, $n$ the refractive index of the gas under
standard conditions, and $\rho$ the depolarization factor.
The refractive index is usually wavelength dependent, and \pymiedap can 
compute the refractive indices of N$_2$, air, CO$_2$, H$_2$ and He,
using dispersion formulae that are valid across visible and near--IR wavelengths
\citep{Peck1966,Ciddor1996,Bideau-Mehu1973,Peck1977,Mansfield1969}. 
Users can also provide their own values for $n$.

Aerosol particles are small particles that are suspended in 
the atmospheric gas. \pymiedap considers cloud particles (relatively large 
particles with relatively high volume number densities) as any other type
of aerosol particle. The influence of aerosol particles on the transfer of
radiation through an atmospheric layer depends on the layer's
aerosol extinction optical thickness $b^{\rm a}$, their single scattering
albedo $a^{\rm a}$ and their single scattering matrix ${\bf F}^{\rm a}$.

The \pymiedap user can specify $b^{\rm a}$ at all required wavelengths,
or provide a value for $N^{\rm a}$, the layer's aerosol column number density
(in $\mu$m$^{-2}$). In the latter case, \pymiedap computes $b^{\rm a}$ from 
$N^{\rm a}$ and the aerosol extinction cross section $\sigma^{\rm a}$,
as follows
\begin{equation}
   b^{\rm a}(\lambda) = \sigma^{\rm a}(\lambda) \hspace*{0.1cm} N^{\rm a}.
\label{eq_ba}   
\end{equation}
Given the microphysical properties of the aerosol particles, 
\pymiedap uses a Mie--algorithm \citep[][]{1984A&A...131..237D} to 
compute $\sigma^{\rm a}$, and through that $b^{\rm a}$, 
for every $\lambda$, assuming that the particles
are homogeneous and spherical. The microphysical properties to be specified by
the user are the particle size--distribution \citep[see][]{1984A&A...131..237D} 
and the refractive index. For layered spherical particles, \pymiedap uses
an adaptation of the algorithm presented by \citet{Bohren1983}. For these types
of particles, the user specifies the refractive indices of the core and the shell,
and the core radius as a fraction of the particle radius.

Using the Mie-algorithm or the adapted algorithm for layered spheres, 
\pymiedap also computes the aerosol single scattering matrix ${\bf F}^{\rm a}$,
which has the following form  
\begin{equation}
   {\bf F}^{\rm a}(\Theta,\lambda) =
          \left[ \begin{array}{cccc}
     A^{\rm a}_1(\Theta,\lambda) & B^{\rm a}_1(\Theta,\lambda) & 0 & 0 \\
     B^{\rm a}_1(\Theta,\lambda) & A^{\rm a}_2(\Theta,\lambda) & 0 & 0 \\
     0 & 0 & A^{\rm a}_3(\Theta,\lambda) & B^{\rm a}_2(\Theta,\lambda) \\
     0 & 0 & -B^{\rm a}_2(\Theta,\lambda) & A^{\rm a}_4(\Theta,\lambda)
          \end{array} \right],
\label{eq_scatmata}
\end{equation} 
and that they are normalized like scattering matrix ${\bf F}^{\rm m}$
(Eq.~\ref{eq_scatmat}). This matrix form holds for spherical particles,
for particles with a plane of symmetry in random orientation, and for particles
that are asymmetric and randomly oriented, while half of the particles are
mirror images of the other half \citep[see][]{1974SSRv...16..527H}.
Rather than the scattering matrix elements themselves, \pymiedap uses 
the coefficients of their expansion into generalized spherical functions
\citep[][]{1984A&A...131..237D}.

Obviously, in nature not all aerosol particles are spherical, and while
\pymiedap cannot compute the expansion coefficients that describe the 
scattering of light by non--spherical particles, it can use them when the user
provides them. Examples of sources of expansion
coefficients of non--spherical particles are those derived from measured
matrix elements, such as those in the Amsterdam--Granada Light Scattering 
Database \citep[][]{2012JQSRT.113..565M}
For differently shaped particles, such as spheroids or ice crystals,
various algorithms have been developed to calculate scattering matrix 
elements, such as the T--matrix method 
\citep[see][]{2002sael.book.....M} and the ADDA--method
\citep[][]{2011JQSRT.112.2234Y}. Expansion coefficients 
derived from those matrix elements could be imported into \pymiedapx.
 
Finally, \pymiedap computes the atmospheric layer's total optical 
thickness $b$ at wavelength $\lambda$ as 
\begin{equation}
   b(\lambda)= b^{\rm m}_{\rm sca}(\lambda) + b^{\rm m}_{\rm abs}(\lambda) +
               b^{\rm a}_{\rm sca}(\lambda) + b^{\rm a}_{\rm abs}(\lambda)
             = b^{\rm m}(\lambda) + b^{\rm a}(\lambda),
\label{eq_btot}   
\end{equation}
the layer's single scattering albedo $a$ as
\begin{equation}
   a(\lambda) = \frac{b^{\rm m}_{\rm sca}(\lambda) + b^{\rm a}_{\rm sca}(\lambda)}
            {b(\lambda)},
\label{eq_atot}            
\end{equation}
and its single scattering matrix ${\bf F}$ as
\begin{equation}
   {\bf F}(\Theta,\lambda) = 
       \frac{b^{\rm m}_{\rm sca}(\lambda) \hspace*{0.1cm} {\bf F}^{\rm m}(\Theta,\lambda)
           + b^{\rm a}_{\rm sca}(\lambda) \hspace*{0.1cm} {\bf F}^{\rm a}(\Theta,\lambda)}
            {b^{\rm m}_{\rm sca}(\lambda) + b^{\rm a}_{\rm sca}(\lambda)}.
\label{eq_ftot}            
\end{equation}
Note that if more than one aerosol type (size distribution, shape, refractive index) 
is used in an atmospheric layer, 
\pymiedap computes the extinction optical 
thickness, single scattering albedo and single scattering matrix of the mixture 
of aerosol particles using equations similar to Eqs.~\ref{eq_btot}--\ref{eq_ftot}
before combining the aerosol optical properties with those of the gaseous molecules.

The values for $b$, $a$, and ${\bf F}$ for every atmospheric layer 
are fed into the adding--doubling radiative transfer algorithm, together
with the reflection properties of the surface.

\subsection{The model surface}
\label{sect4.2}

Unless it is pitch--black, a planetary surface will reflect incident direct 
light, i.e.\ the unscattered light from the sun or star, and, if there is an 
atmosphere above the surface, the incident diffuse light, i.e.\ the light from 
the sun or star that has been scattered and that emerges from the bottom of the 
atmosphere. 
The surface albedo $a_{\rm s}$ indicates the fraction of all incident light that 
is reflected back up. This albedo ranges from 0.0 (all incident light is absorbed) 
to 1.0 (all incident light is reflected).

In \pymiedapx, the surface reflection is defined through a reflection matrix. 
In the current version of \pymiedapx, the surface reflection is Lambertian:
it reflects light isotropically and completely depolarized. The (1,1)-element 
of the reflection matrix of a Lambertian surface equals $a_{\rm s}$ and 
is thus independent of the illumination and viewing geometries, while all 
other matrix elements equal zero.


\section{The radiative transfer algorithm}
\label{dap}

As described in Eq.~\ref{eq_reflvector2}, to calculate Stokes 
vector ${\bf I}(\mu,\mu_0,\phi-\phi_0)$ of light that is locally reflected by
a model planet, we have to compute ${\bf R}_1$, the first column of the local 
planetary reflection matrix. The computation of vector ${\bf R}_1$ includes
all orders of scattering in the planetary atmosphere and reflection by 
the surface (if $a_{\rm s} > 0.0$).
\pymiedap computes ${\bf R}_1$, although not directly. Instead, \pymiedap
produces ASCII-files (see Appendix \ref{fourier-files})
that contain the coefficients of the Fourier expansion 
of ${\bf R}_1$ for various combinations of $\mu$ and $\mu_0$. These files,
that are stored in a database for repeated use,
are accessed to compute ${\bf R}_1$ for the required combination of 
$(\mu,\mu_0,\phi-\phi_0)$. The expansion is described in Sect.~\ref{sect5.1},
and the subsequent application for the required geometry in 
Sect.~\ref{sect5.2}.

\subsection{The Fourier expansion of the reflection matrix}
\label{sect5.1}

Equation~28 in \citet{1987A&A...183..371D} shows how a matrix
such as the planetary reflection matrix ${\bf R}$ can be expanded in a 
Fourier series. Because we only need the first column of this matrix,
we can rewrite this expansion as follows
\begin{multline}
   {\bf R}_1(\mu,\mu_0,\phi-\phi_0,\lambda) = 
        {\bf B}^{+0}(\phi-\phi_0) \hs {\bf R}^0_1(\mu,\mu_0,\lambda) \hs + \\
        2 \sum_{m=1}^{M} {\bf B}^{+m}(\phi-\phi_0) \hs {\bf R}^m_1(\mu,\mu_0,\lambda),
\label{eq_fou2}
\end{multline}
where ${\bf R}^m_1$ is the first column of the $m$th Fourier coefficient 
matrix ${\bf R}^m$ ($0 \leq m \leq M$). 
The series is summed up till and including coefficient number $M$, the value
of which is determined by the accuracy of the adding--doubling 
radiative transfer calculations \citep[see][]{1987A&A...183..371D}.
Matrices ${\bf B}^{+m}$ and ${\bf B}^{-m}$ have zero's everywhere 
except on the diagonal:
\begin{eqnarray}
   {\bf B}^{+m}(\phi) & = & {\rm diag}(\cos m\phi, \cos m\phi, \sin m\phi, \sin m\phi), \\
   {\bf B}^{-m}(\phi) & = & {\rm diag}(-\sin m\phi, -\sin m\phi, \cos m\phi, \cos m\phi).
\label{eq_B}
\end{eqnarray}
An obvious advantage of using the Fourier coefficients vectors ${\bf R}^m_1$ instead
of ${\bf R}_1$ itself, is that they are independent of the azimuthal
angle difference $\phi-\phi_0$.
Combining Eqs.~\ref{eq_reflvector2} and ~\ref{eq_fou2}-\ref{eq_B}, 
the elements of vector ${\bf I}$ describing the
light that is locally reflected by a planet are obtained through
\begin{multline}
   I(\mu,\mu_0,\phi-\phi_0,\lambda) / \mu_0 F_0(\lambda) = \\ 
         R_{11}^0(\mu,\mu_0,\lambda) \hs + 
         2 \sum_{m=1}^{M} \cos m(\phi-\phi_0) \hs 
         R^m_{11}(\mu,\mu_0,\lambda), 
\label{eq1}
\end{multline}
\begin{multline}
   Q(\mu,\mu_0,\phi-\phi_0,\lambda) / \mu_0 F_0(\lambda)  = \\ 
         R_{21}^0(\mu,\mu_0,\lambda) \hs +
         2 \sum_{m=1}^{M} \cos m(\phi-\phi_0) \hs
         R^m_{21}(\mu,\mu_0,\lambda), 
\end{multline}
\begin{multline}
   U(\mu,\mu_0,\phi-\phi_0,\lambda) / \mu_0 F_0(\lambda) = 
         2 \sum_{m=1}^{M} \sin m(\phi-\phi_0) \hs 
         R^m_{31}(\mu,\mu_0,\lambda), 
\end{multline}
\begin{multline}
   V(\mu,\mu_0,\phi-\phi_0,\lambda) / \mu_0 F_0(\lambda) =
         2 \sum_{m=1}^{M} \sin m(\phi-\phi_0) \hs
         R^m_{41}(\mu,\mu_0,\lambda), 
\label{eq4}
\end{multline}
with the subscripts $11$, $21$, $31$, and $41$ denoting the 1st, 2nd, 3rd, and 
4th element of the column vectors ${\bf R}^0_1$ and ${\bf R}^m_1$, respectively.
For a given model planet, the Fourier file contains $R^m_{11}$, $R^m_{21}$,
$R^m_{31}$, and $R^m_{41}$ for $m=0$ to~$M$ 
for various combinations of $\mu$ and $\mu_0$ (see Sect.~\ref{sect5.2}).

We calculate $R^m_{11}$, $R^m_{21}$, $R^m_{31}$, and $R^m_{41}$ 
using the accurate and efficient adding--doubling radiative transfer 
algorithm as described in \citet{1987A&A...183..371D}. 
This algorithm includes all orders of scattering, and it fully includes 
linear and circular polarization for all orders.


\subsection{Gaussian abscissae}
\label{sect5.2}

The values of $\mu$ and $\mu_0$ at which the Fourier coefficients are provided, 
equal the Gaussian abscissae that are used in the adding--doubling algorithm 
\citep[][]{1987A&A...183..371D} for the Gauss--Legendre integrations over 
all scattering directions. 
For example, if 12~abscissae are
used for the integrations, we provide the coefficients $R^m_{z1}$ (with $z$
equal to 1, 2, 3, or 4) at these 12 values of $\mu_0$ and at the same 12~values of
$\mu$, thus at a total of 144 combinations of illumination ($\mu_0$) and viewing
($\mu$) geometries. 

The number of Gaussian abscissae 
that is required to reach a given accuracy with the radiative transfer computations
depends strongly on the single scattering
properties of atmospheric aerosol particles. In particular, if the
scattered total and polarized fluxes vary strongly with the
single scattering angle (typically when the particles are large with respect to the 
wavelength $\lambda$, see e.g. \cite{1974SSRv...16..527H} for sample figures), 
more abscissae are needed than when they vary smoothly.
The required number of abscissae depends also on the illumination and viewing geometries,
for example, large solar zenith angles and/or viewing angles usually require 
more abscissae than small angles.
We choose the number of abscissae in the database files such that the coefficients will
give accurate results for a large range of combinations of $\mu$ and $\mu_0$.

The expansion coefficients provided in a Fourier coefficients 
file can be used directly
to evaluate Eqs.~\ref{eq1}-\ref{eq4} at one of the available combinations of Gaussian 
abscissae $\mu$ and $\mu_0$, and for an arbitrary, user defined, 
value of $\phi-\phi_0$.
Fourier coefficients at values of $\mu$ and/or $\mu_0$ that do not coincide
with Gaussian abscissae can be obtained by interpolation.

To avoid having to extrapolate to obtain
results at the often used values of $\mu$ and/or $\mu_0$ equal to 1.0 
(i.e. $\theta$, $\theta_0=0^\circ$),
which are not part of any set of Gaussian abscissae 
(that have values larger than 0.0 and smaller than 1.0),  
we have included $\mu_0$, $\mu_0=1.0$ as so--called supplemented
$\mu$-values (see Sect.~5 of \citet{1987A&A...183..371D}). The adding--doubling 
algorithm calculates the Fourier coefficients at these supplemented values
as if they were Gaussian abscissae. 
Thus, if we use $M$~Gaussian abscissae, 
the Fourier coefficients are provided at $M+1$ values of $\mu$ and $\mu_0$
(thus at $(M+1)^2$ combinations of $\mu$ and $\mu_0$).

\pymiedap separates the computation and storage of the Fourier coefficients from
the computations of the locally reflected light (Eqs.~\ref{eq1}-\ref{eq4}),
and it can indeed skip the Fourier coefficients computation, and instead
use a previously computed Fourier file to compute the locally reflected light.
The advantage of this is that time is saved, because depending on 
the composition and structure of the model atmosphere, the Fourier computations 
can take a significant amount of computing time.


\section{Horizontally inhomogeneous planets}
\label{geometry-masks}

Because \pymiedap is pixel--based, it can be applied to 
horizontally inhomogeneous planets, i.e. planets with horizontal variations in 
their atmosphere and/or surface properties. The user can define such horizontal
inhomogeneities by using a so--called mask, in which pixels are assigned a 
value corresponding to a specific atmosphere--surface model combination, e.g.\
'0' for model combination 1, '1' for the model combination 2, ... 
When \pymiedap computes the locally reflected light for a given pixel, 
it will do so using the Fourier coefficients file for the model 
associated with that pixel.
A pixel mask can be phase angle dependent, i.e. a different pattern can be defined
for each phase angle, e.g. to simulate a rotating planet.

\begin{figure}[h]
    \centering
    \begin{subfigure}[t]{0.49\linewidth}
        \caption{A sub-solar cloud}
        \vspace{-1ex}
        \includegraphics[width=\linewidth]{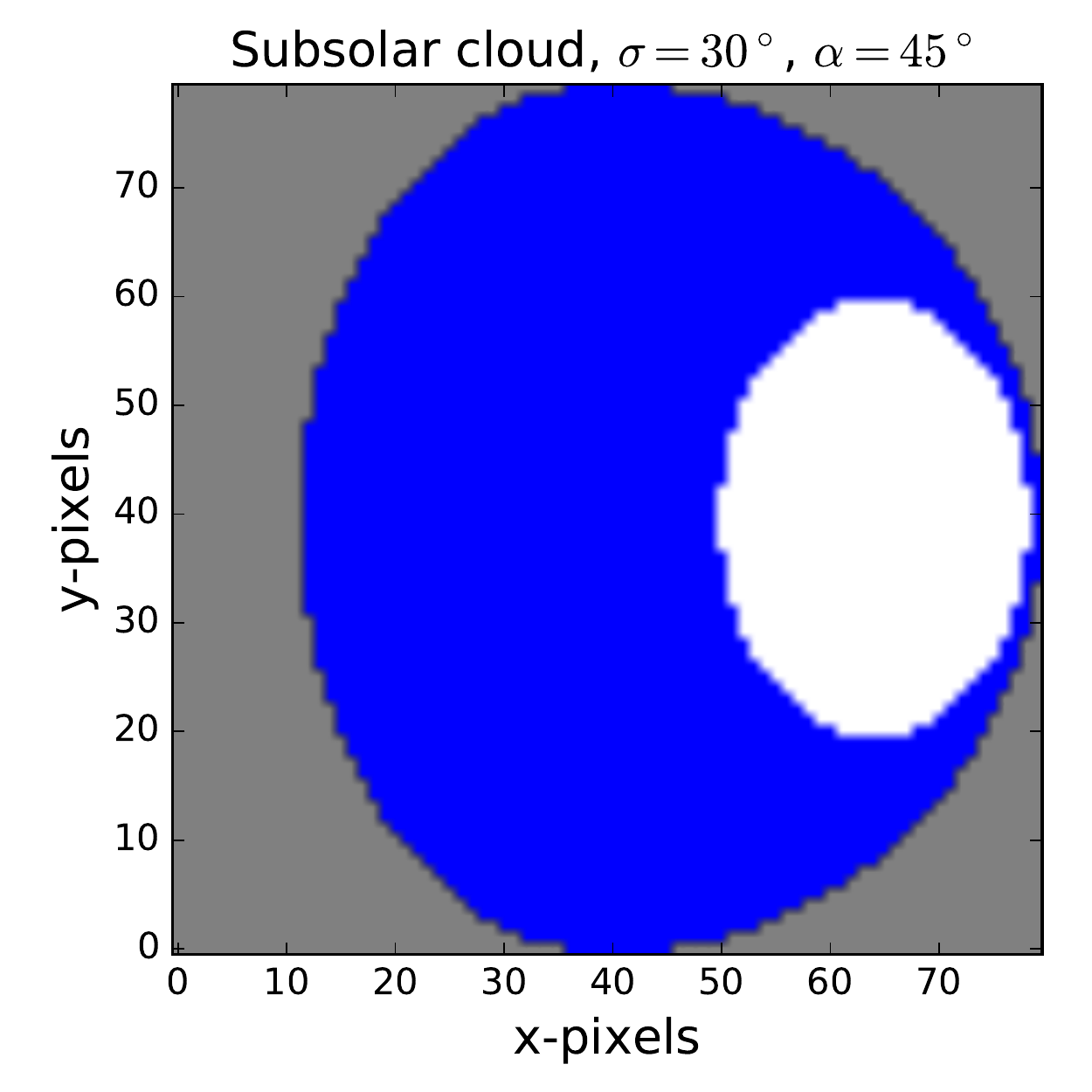}
    \end{subfigure}
    \begin{subfigure}[t]{0.49\linewidth}
        \caption{Polar cusps}
        \vspace{-1ex}
        \includegraphics[width=\linewidth]{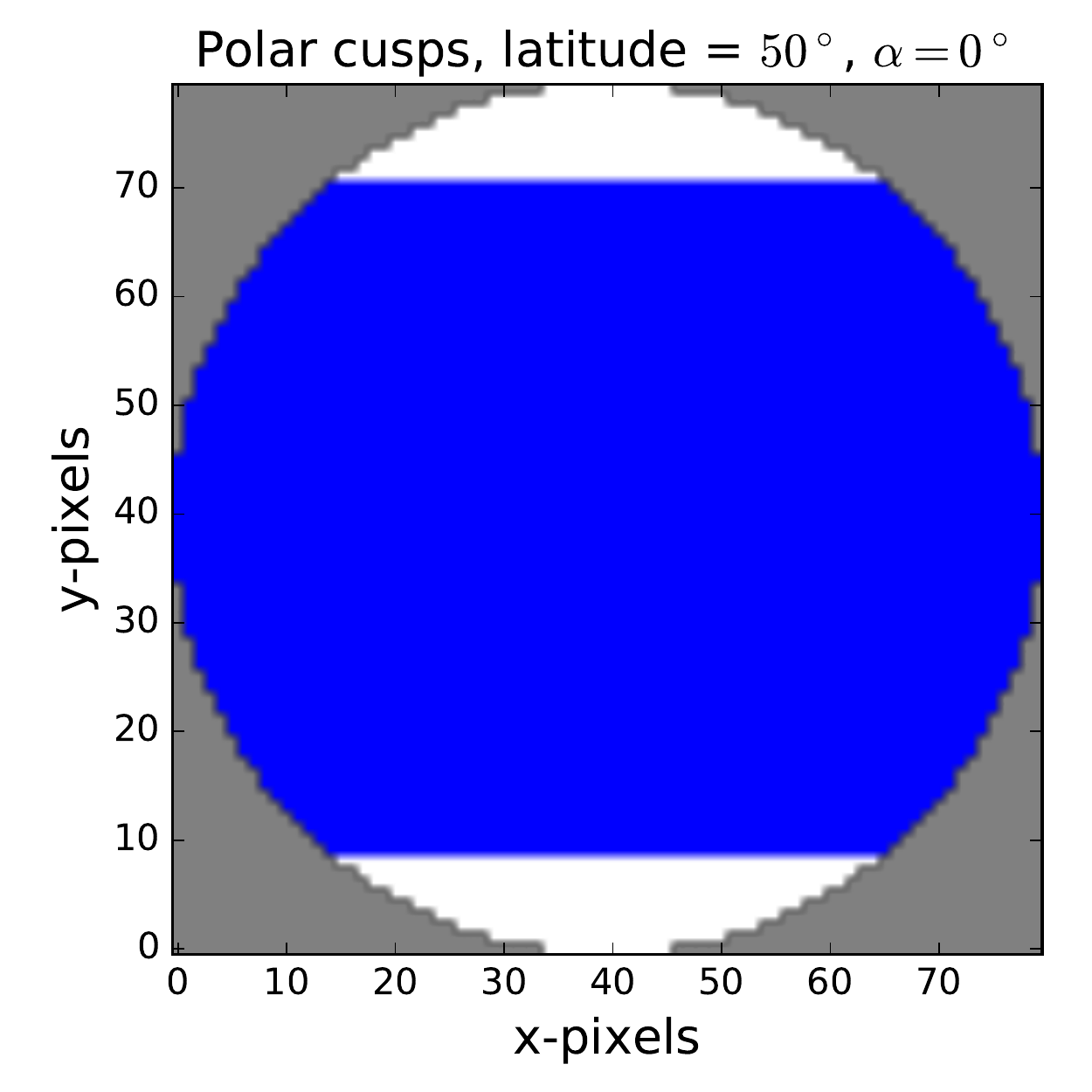}
    \end{subfigure}
    \vspace*{0.2cm}
    \begin{subfigure}[t]{0.49\linewidth}
        \caption{Latitudinal bands}
        \vspace{-1ex}
        \includegraphics[width=\linewidth]{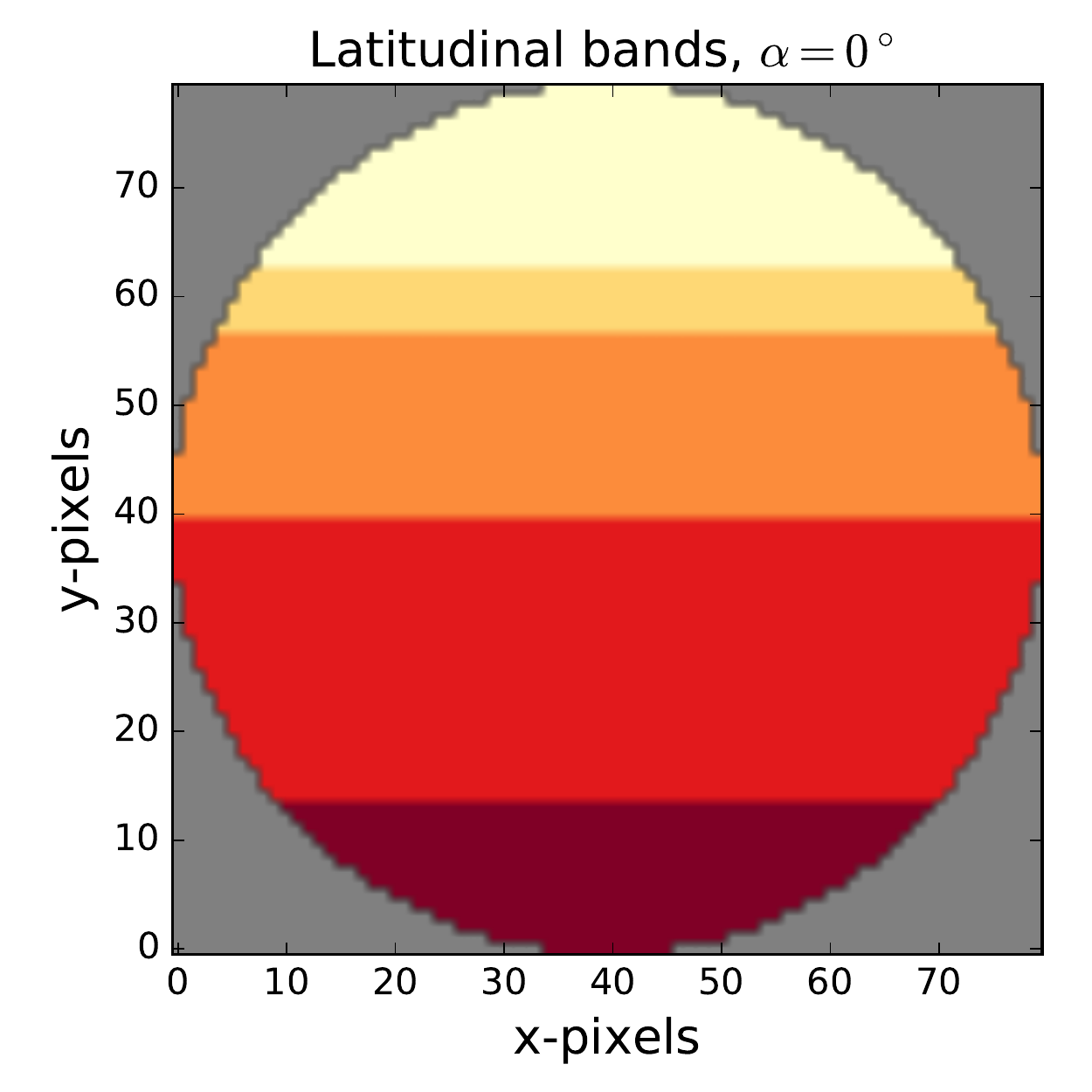}
    \end{subfigure}
    \begin{subfigure}[t]{0.49\linewidth}
        \caption{Patchy clouds}
        \vspace{-1ex}
        \includegraphics[width=\linewidth]{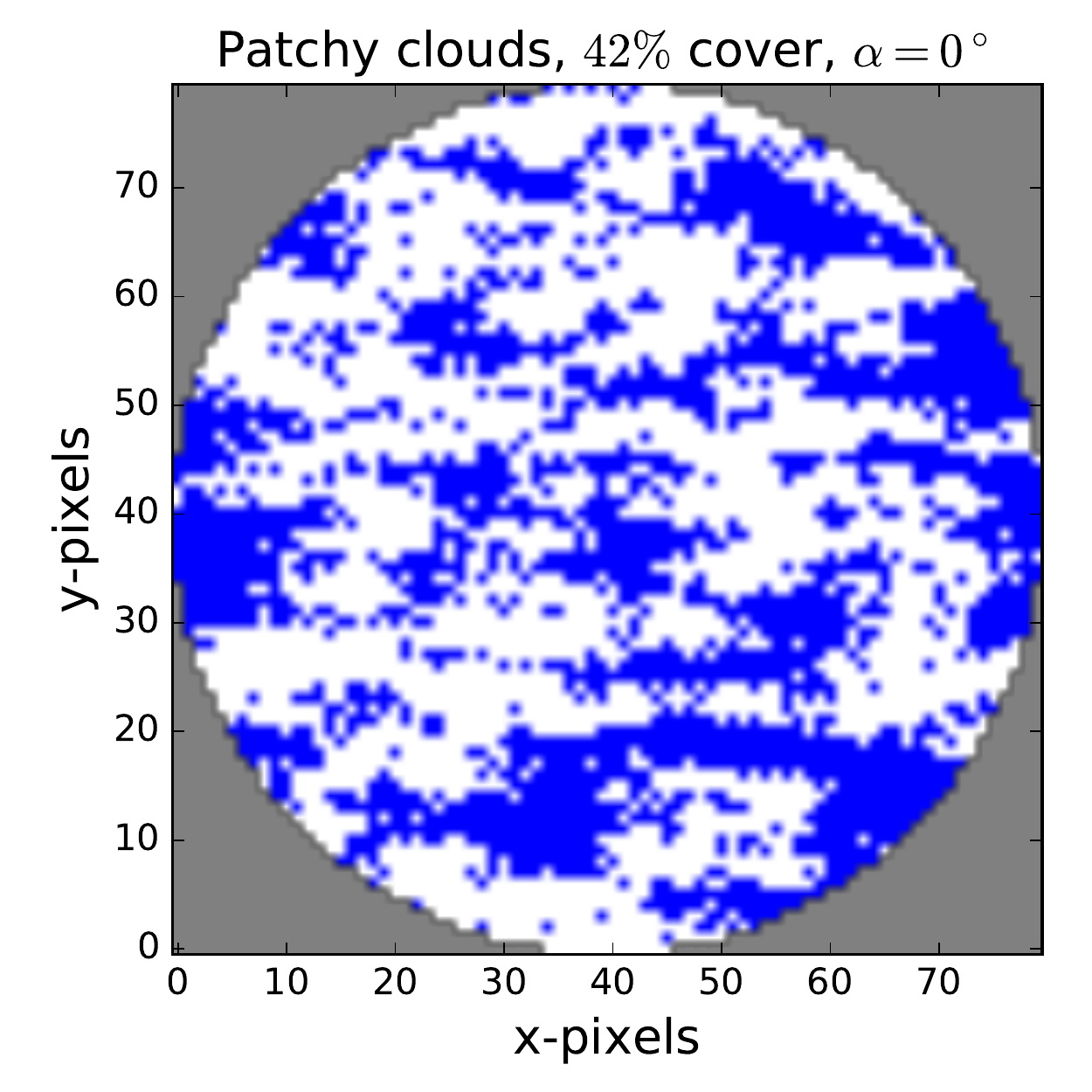}
    \end{subfigure}
    \caption{Examples of four types of cloud cover:
             (a) Sub-solar clouds with an angular width $\sigma_{\rm c}$ of $30^\circ$
             at $\alpha=45^\circ$;
             (b) Polar cusps for a threshold latitude $L_{\rm t}$ of 50$^\circ$
             at $\alpha=0^\circ$;
             (c) Latitudinal bands with borders at $-90^\circ$, $-40^\circ$,
             $0^\circ$, $25^\circ$, and $35^\circ$, $90^\circ$;
             (d) Patchy clouds for a cloud fraction $F_{\rm c}= 0.42$ at 
             $\alpha=0^\circ$.
             In all figures, $N_{\rm eq}=80$.}
\label{fig:cover_types}
\end{figure}

Common horizontal inhomogeneities on planets are clouds. \pymiedap\ has 
the following 4 different types of cloud coverage masks built in 
(see Fig.~\ref{fig:cover_types}): 
sub-solar clouds, polar cusps, latitudinal bands, and patchy clouds. \\
    
\noindent
{\em Sub-solar clouds} are thought to be relevant for tidally--locked planets,
such as exoplanets in tight orbits around their parent star \citep{Yang2013}. 
For these clouds, the pixel grid on the planetary disk is filled such that 
the region where the local solar zenith angle $\theta_0$ is smaller than the
user--defined angle $\sigma_{\rm c}$ is assigned one atmosphere-surface 
model combination, and the other pixels another. This mask can also be
used to model a sub-solar ocean 
\citep[also referred to as "eyeball planet";][]{Turbet2016,Pierrehumbert2011}. \\

\noindent
{\em Polar--cusps} are clouds that form where the daily averaged incident 
solar or stellar flux dips below a certain threshold. For these clouds, the
pixels located poleward of the user--defined latitude $L_{\rm t}$ on the planet
are assigned one model combination, and the other pixels another (the planet's
equator is assumed to be in the middle of the planetary disk). This mask would
also be useful to model polar hazes. \\

\noindent
{\em Latitudinal bands} are bands of clouds covering ranges of latitudes.
For these clouds, the user provides an array of latitudes that border
the different atmosphere-surface models (the planet's
equator is assumed to be in the middle of the planetary disk).
Such a mask can be useful for planets with belts and zones, or 
to simulate planets with latitudinal variations.  \\
    
\begin{figure}
\centering
\includegraphics[width=0.7\linewidth]{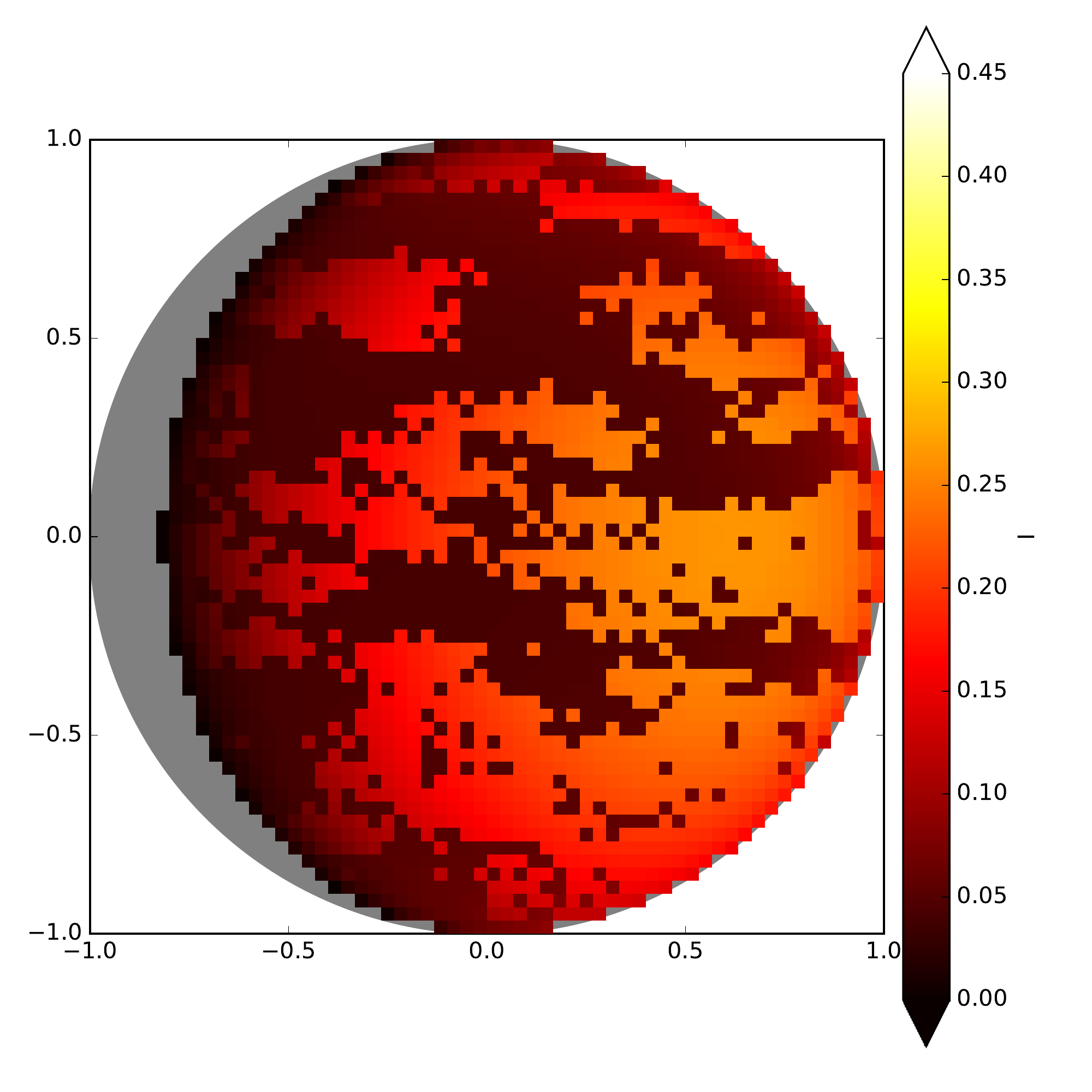}\\
\includegraphics[width=0.7\linewidth]{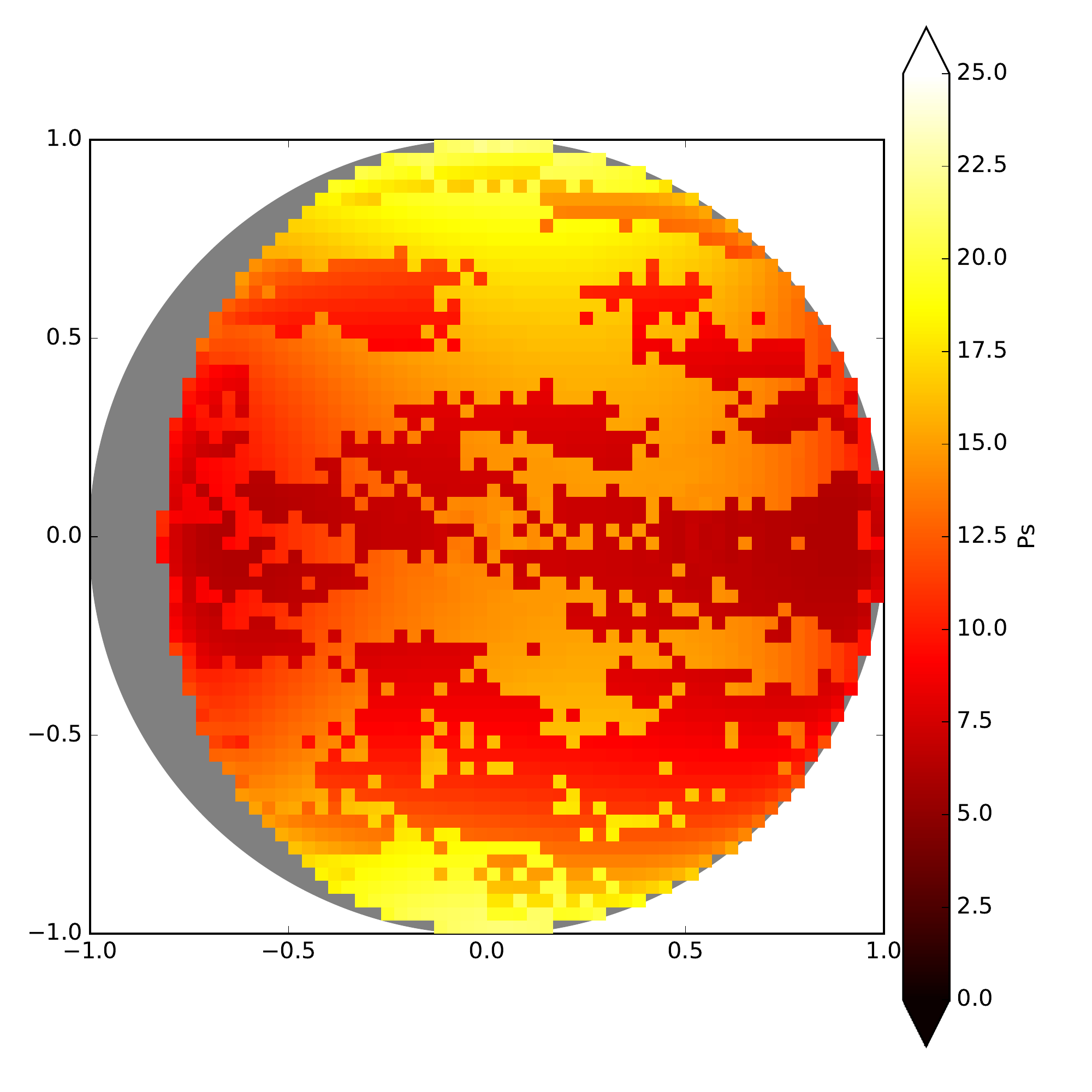}\\
\includegraphics[width=0.7\linewidth]{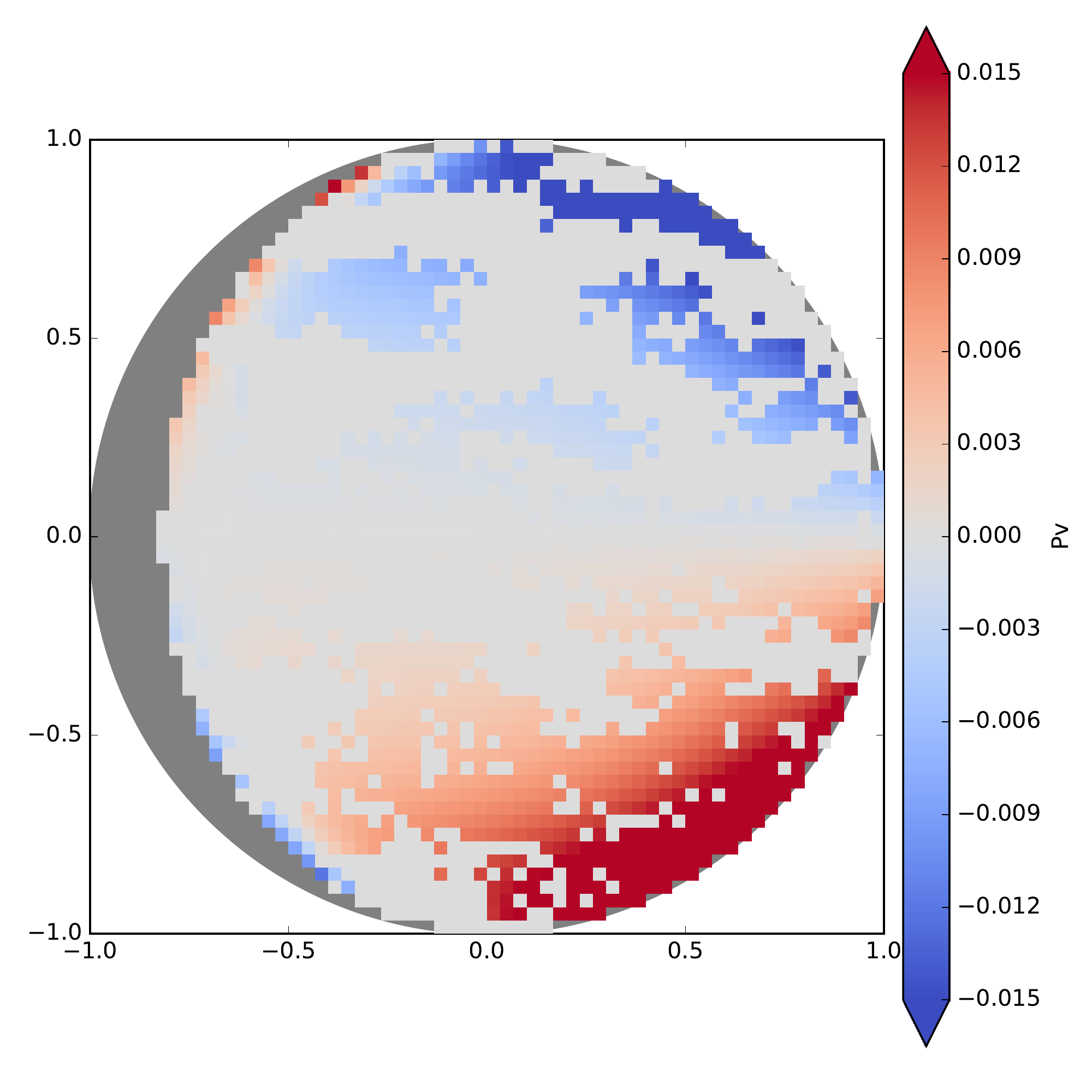}
\caption{Computed disk--resolved locally reflected fluxes $I$ (top), 
         $P_{\rm s}$ (middle, in percents) and $P_{\rm c}$ (bottom, in percents) at $\alpha=35^\circ$ 
         and $\lambda=0.55~\mu$m, for a patchy cloud cover with 
         $F_{\rm c}=0.50$. The cloud--free pixels have a pure gaseous (CO$_2$) 
         model atmosphere with $b=7.0$.
         The cloudy pixels contain aerosol as described by \citet{Stam2006} 
         (we let the model D type aerosol pose as cloud particles).  
         The disk--integrated values are $F=0.32$, $P_{\rm s}=0.10$, and
         $P_{\rm c}= 1.6 \cdot 10^{-5}$. For all figures, $N_{eq}=60$.
         }
\label{fig_gasgiants}
\end{figure}

\noindent
{\em Patchy clouds} are distributed across the planetary disk. 
They are described by $F_{\rm c}$, the fraction of all pixels on the 
disk that are cloudy, and the spatial distribution of these cloudy pixels.
This mask can accept $N$ atmosphere-surface models, 
each with its own cloud coverage fraction $F_{\rm c,i}$, 
with $i=0,\dots,N-1$. 
Patchy cloud patterns are generated by drawing 50 values from a 
2D--Gaussian distribution centred on a randomly chosen location 
within the pixel grid. The covariance matrix is given by 
\begin{equation}
   \Sigma = n_{\rm pix}
   \left[
   \begin{array}{c c}
   x_{\rm scale} & 0 \\
   0 & y_{\rm scale} \\
   \end{array}
   \right]
\end{equation}
where $x_{\rm scale}$ and $y_{\rm scale}$ are used to fine--tune the shapes
of the cloud patches along the north--south and east--west axes.
\pymiedap uses $x_{\rm scale}= 0.1$ and $y_{\rm scale}= 0.01$ as nominal 
values in order to generate clouds with a zonal--oriented
pattern similar to that observed on Earth. 
Cloud patches are generated on the planetary disk until the specified value of 
$F_{\rm c,i}$ is reached, the overall cover of all types of patches being 1.
The cloud fraction $F_{\rm c}$ is defined at $\alpha=0^\circ$, because climatologically,
the planetary--wide cloud coverage is more relevant than the coverage seen by 
an observer. The cloud fraction observed at $\alpha > 0^\circ$
can thus differ from the specified value of $F_{\rm c}$.
An illustration of this cloud distribution is given in figure \ref{fig_gasgiants}, which shows the disk-resolved simulations of flux, linear and circular polarisation for $50\%$ cloud cover, as computed by \pymiedapx.

The disk--integrated signal of a planet covered by patchy clouds will depend on 
the position of the cloudy pixels on the disk. To capture this variability, 
the user can choose to draw several patterns randomly at each phase angle. 
\pymiedap will return the average and standard deviations of the values of 
$I$, $Q$, $U$, and $V$ over all patterns. It can also store the values for each 
pattern, providing the user insight into the variability.

An example of the variability computed using \pymiedap can be found in 
\citet{Rossi2017} where it was used to generate the disk--integrated 
signals of Earth--like exoplanets with varying types and amount of 
coverage by liquid water clouds. 
Thanks to the use of Fourier coefficients, only a limited number of model
computations were necessary: clear sky and cloudy case with different
cloud--top altitudes. Furthermore, because
the Fourier files allow for computation of the reflected Stokes vector of light
for any geometry, it was possible to generate the Stokes vector of each pixel
for the clear and cloudy cases, and then apply masks on these grids of pixels
to obtain the desired cloud pattern. The variability due to patchy cloud cover
could be simulated by simply using 300 patterns that were averaged.

Another example of use of patchy cloud masks in \pymiedap can be found in 
\citet{Fauchez2017} where disk--integrated signals of exoplanets with patchy 
clouds were computed to investigate the effect of such clouds on the 
spectral signature of the O$_2$ A-absorption band in the flux and polarization
of reflected starlight.


\section{Benchmark results}
\label{benchmark}

Here, we will compare results of \pymiedap against (published) results obtained 
with other codes. This comparison allows an assessment
of the accuracy of \pymiedapx's approach using computed Fourier coefficients files, 
and our results allow \pymiedap users to check their \pymiedap installation 
and understanding of the input and output files. 

\subsection{Locally reflected light}
\label{sect7.1}

We compare our results for locally reflected light with those presented 
in Tables~5, 6, 9 and 10 of \citet{1987A&A...183..371D}.
We use the same adding-doubling algorithm as \citet{1987A&A...183..371D},
with the same accuracy, i.e.\ 10$^{-6}$.
However, while \citet{1987A&A...183..371D} compute the reflected Stokes
vectors at precisely the specified values of $\theta_0$ and $\theta$,
we use a Fourier coefficients file (with $\theta = \theta_0 = 0^\circ$ 
as supplemented Gaussian abscissae), combined with spline interpolation
\citet[with the algorithm from][]{1992nrfa.book.....P} to obtain the
reflected Stokes vectors at the same geometries.

Two model atmosphere-surface combinations are considered in 
\citet{1987A&A...183..371D}: model 1, with a single layer atmosphere 
containing only haze droplets, bounded below by a black surface, and 
model 2, with an upper atmospheric layer containing only
gas and a second, lower layer containing a mixture
of gas and haze droplets, bounded below by a 
Lambertian reflecting surface with albedo $A_{\rm s}$ of 0.1.
The molecular depolarization factor $\rho$ is 0.0279. The 
haze particles in both atmospheres are water--haze L particles 
\citep[][]{1969esos.book.....D}, with their optical
properties calculated at $\lambda= 0.7~\mu$m
\citep[for the single scattering expansion coefficients, see][]{1984A&A...131..237D}.
For model 1, $b^{\rm a} = b^{\rm a}_{\rm sca} = 1.0$.
For model~2, $b^{\rm m} = b^{\rm m}_{\rm sca}= 0.1$ in each layer,
and in the lower layer, $b^{\rm a} = b^{\rm a}_{\rm sca}= 0.4$.
The incident flux $\pi F_0$ equals $\pi$ ($F_0$ is thus 1.0).

Table~\ref{benchmark-modA} shows the Stokes vector elements of the 
locally reflected light for model~1 from \citet{1987A&A...183..371D} and
calculated using \pymiedap and 40 Gaussian abscissae ($N_{\rm G}= 40$).
Table~\ref{benchmark-modB} shows the results for model~2.
As can be seen, \pymiedapx's pre-calculated 
Fourier coefficients combined with spline interpolation 
yields accurate results for both models.
Note that when $\mu= 1.0$, i.e.\ the supplemented 
Gaussian abscissa in our Fourier files (cf.\ Sect.~\ref{sect5.2}),
we only have to interpolation between Fourier coefficients
for $\mu_0$, not for $\mu$.

\subsection{Disk-integrated reflected starlight}
\label{sect-disk-int}

\begin{figure}[h]
\centering
\includegraphics[width=\linewidth]{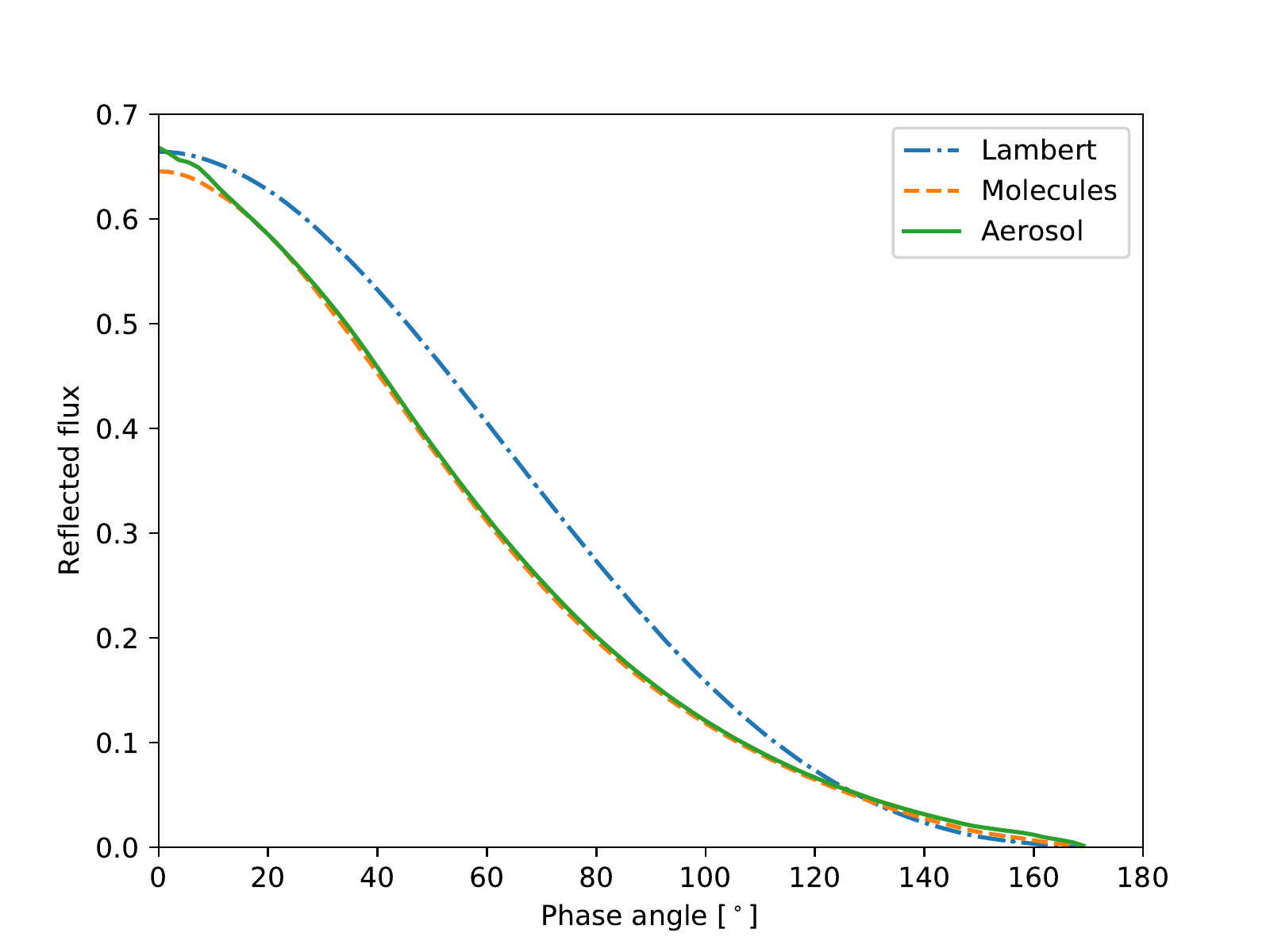}\\
\includegraphics[width=\linewidth]{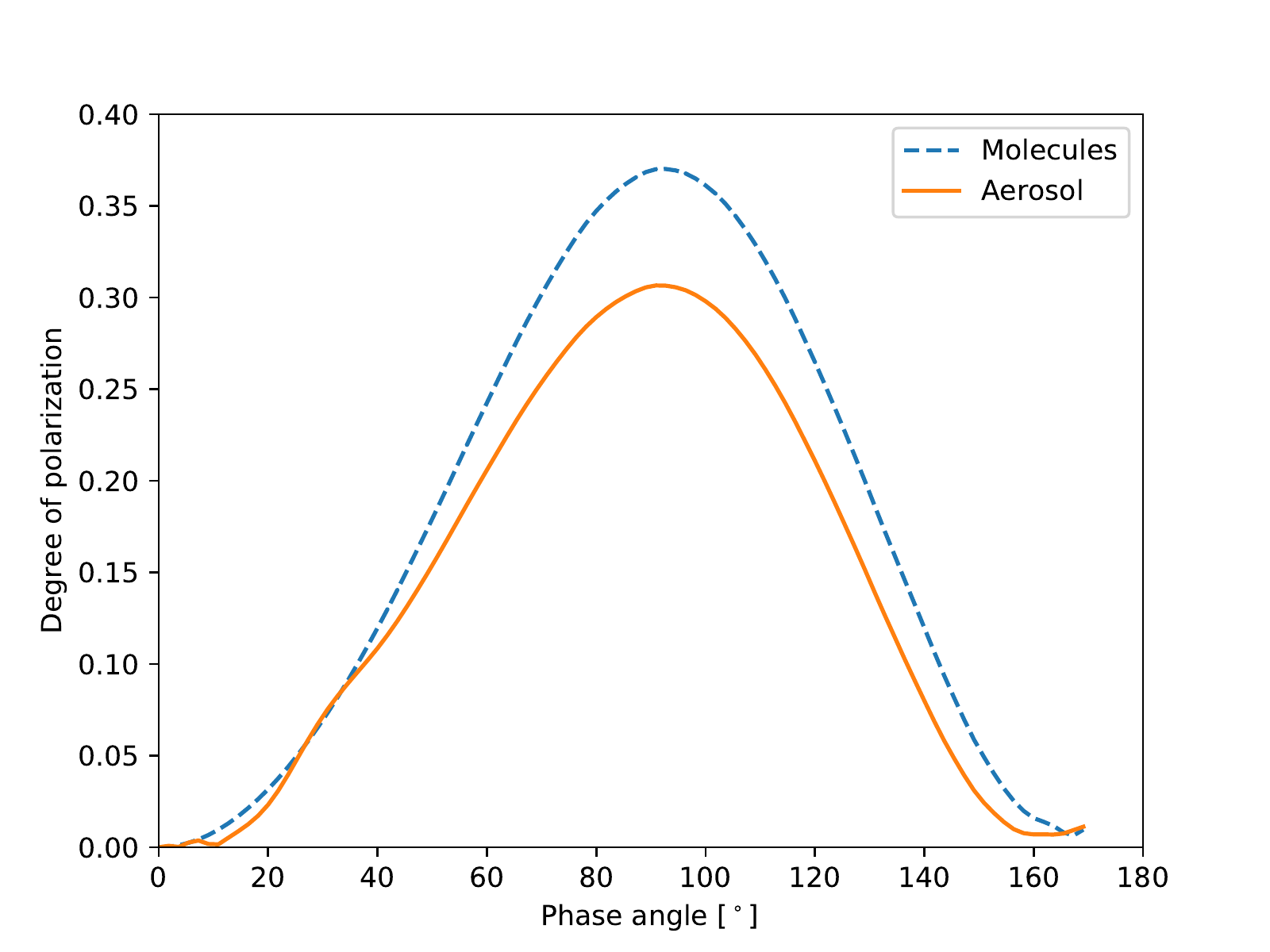}
\caption{Numerical simulations of disk--integrated reflected fluxes (top) 
         and the degree of linear polarization (bottom) as functions of 
         $\alpha$ for different model planets. The flux curves are 
         normalized such that they equal the planet's geometric albedo
         $A_{\rm G}$ at $\alpha=0^\circ$. For all curves, $N_{eq} = 20$.
         Dot-dashed line (only flux): a Lambertian reflecting planet with a 
         surface albedo $a_{\rm s}=1.0$.
         Dashed line: a planet with a gaseous atmosphere with 
         $b_{\rm sca}^{\rm m}=5.75$, $\rho=0.02$, and $a_{\rm s}=0.0$.
         Solid line: the same planet with model~D aerosol (see text) added.
         }
\label{fig_diskint}
\end{figure}

There are no disk--integrated Stokes parameters in \citet{1987A&A...183..371D}.
We therefore first compare the disk--integrated reflected total flux as 
computed using \pymiedap against the 
analytical expression for the phase function of a Lambertian reflecting, 
spherical planet with a surface albedo $a_{\rm s}$, i.e.\ 
\citep[see][]{1980vandeHulst}
\begin{equation}
   \psi(\alpha)= \frac{2}{3\pi} a_{\rm s} \left( \sin \alpha + \pi \cos \alpha - 
              \alpha \cos \alpha \right).
\label{eq_lam}
\end{equation}
Figure~\ref{fig_diskint} shows $\psi$ calculated using \pymiedap and 
$a_{\rm s}=1.0$. For $N_{\rm G} \geq 20$, these results are indistinguishable 
from those computed by Eq.~\ref{eq_lam}.
This comparison also shows the validity of calculating reflected disk--integrated 
fluxes under the assumption of a locally plane--parallel atmosphere and/or surface.

To test the accuracy of the computed disk--integrated polarization, 
we compare \pymiedap results against results for two Jupiter--like gas planets 
computed with the same
Fourier expansion coefficients but an integration
method that treats the whole planet as a single scattering particle
\citep[][]{Stam2006} (the latter method is only applicable 
to horizontally homogeneous planets).

The first planet \citep[Sect.~4.1 of][]{Stam2006}
has a purely gaseous atmosphere with $b^{\rm m}_{\rm sca}=5.75$ and 
$b^{\rm m}_{\rm abs}=0.0$, and $\rho=0.02$ (i.e.\ representative for 
H$_2$). The surface albedo $a_{\rm s}=0.0$. 
Figure~\ref{fig_diskint} shows the disk--integrated reflected total 
flux and degree of linear polarization as functions of $\alpha$ 
at $\lambda=0.55~\mu$m.
According to \citet{Stam2006}, the geometric albedo $A_{\rm G}$ of this 
planet is 0.647 and the maximum degree of polarization 0.37 (this value 
is reached at $\alpha=93^\circ$ (note that \citet{Stam2006} use the 
scattering angle $\Theta$ rather than $\alpha$ ($\Theta= 180.0 - \alpha$).
Table~\ref{gas-giant-1} shows both the results of
\citet{Stam2006} (interpolated linearly to obtain the values at the listed 
phase angles) and the results from \pymiedap. 
Comparing the results, it is clear that \pymiedap is very accurate, 
even for a relatively small number of Gaussian abscissae (i.e.\ $N_{\rm G}=20$).

The second planet \citep[Sect.~4.2 of][]{Stam2006}
has the same gaseous atmosphere and black surface as the first, 
but with aerosol particles added. The aerosol optical thickness $b^{\rm a}=3.25$, 
yielding a total atmospheric optical thickness $b$ of 9.0 (at $\lambda=0.55~\mu$m).
The aerosol particles are well--mixed with the gas molecules, 
and have the microphysical properties of the model~D particles of 
\citet[][]{1984A&A...131..237D}.
The disk--integrated reflected flux and degree of linear polarization as
functions of $\alpha$ computed with \pymiedap 
are shown in Fig.~\ref{fig_diskint}.
The geometric albedo $A_{\rm G}$ of this second planet is 0.669 according
to \citet{Stam2006}.
Table~\ref{gas-giant-2} is similar to Table~\ref{gas-giant-1}, except for the
second planet. Because the single scattering scattering matrix elements of model~D
aerosol particles show significant angular structures, in particular in the 
forward and backward scattering directions, more Gaussian abscissae are 
needed (by both disk-integration methods) to achieve accurate results across
the whole phase angle range; we used $N_{\rm G}=50$.

\begin{table}
Stokes parameter $I$ \\
\begin{tabular}{r r r r r}
$\mu_0$ & $\mu$ & $\phi-\phi_0$ & de Haan et al.\ & \pymiedap \\ \hline
0.5 & 0.1 & 0.0 &  1.102690 &  1.102679\\
 0.5 & 0.5 & 0.0 &  0.319430 &  0.319428\\
 0.5 & 1.0 & 0.0 &  0.033033 &  0.033033\\
 0.5 & 0.1 & 30.0 &  0.664140 &  0.664143\\
 0.5 & 0.5 & 30.0 &  0.252090 &  0.252094\\
 0.5 & 1.0 & 30.0 &  0.033033 &  0.033033\\
 0.1 & 0.1 & 0.0 &  2.932140 &  2.932100\\
 0.1 & 0.5 & 0.0 &  0.220540 &  0.220536\\
 0.1 & 1.0 & 0.0 &  0.009287 &  0.009287\\
 0.1 & 0.1 & 30.0 &  0.769100 &  0.769102\\
 0.1 & 0.5 & 30.0 &  0.132828 &  0.132829\\
 0.1 & 1.0 & 30.0 &  0.009287 &  0.009287
 \end{tabular}
 
\vspace*{0.2cm}
Stokes parameter $Q$ \\
\begin{tabular}{r r r r r}
$\mu_0$ & $\mu$ & $\phi-\phi_0$ & de Haan et al.\ & \pymiedap \\ \hline
0.5 & 0.1 & 0.0 &  0.004604 &  0.004604\\
 0.5 & 0.5 & 0.0 & -0.002881 & -0.002880\\
 0.5 & 1.0 & 0.0 & -0.002979 & -0.002979\\
 0.5 & 0.1 & 30.0 &  0.000303 &  0.000303\\
 0.5 & 0.5 & 30.0 & -0.001444 & -0.001443\\
 0.5 & 1.0 & 30.0 & -0.001489 & -0.001489\\
 0.1 & 0.1 & 0.0 &  0.009900 &  0.009900\\
 0.1 & 0.5 & 0.0 &  0.000976 &  0.000977\\
 0.1 & 1.0 & 0.0 & -0.000815 & -0.000815\\
 0.1 & 0.1 & 30.0 & -0.003758 & -0.003758\\
 0.1 & 0.5 & 30.0 &  0.000220 &  0.000220\\
 0.1 & 1.0 & 30.0 & -0.000408 & -0.000408
 \end{tabular}

\vspace*{0.2cm}
Stokes parameter $U$ \\
\begin{tabular}{r r r r r}
$\mu_0$ & $\mu$ & $\phi-\phi_0$ & de Haan et al.\ & \pymiedap \\ \hline
0.5 & 0.1 & 0.0 &  0.000000 &  0.000000\\
 0.5 & 0.5 & 0.0 &  0.000000 &  0.000000\\
 0.5 & 1.0 & 0.0 &  0.000000 &  0.000000\\
 0.5 & 0.1 & 30.0 & -0.002770 & -0.002770\\
 0.5 & 0.5 & 30.0 & -0.004141 & -0.004141\\
 0.5 & 1.0 & 30.0 & -0.002580 & -0.002580\\
 0.1 & 0.1 & 0.0 &  0.000000 &  0.000000\\
 0.1 & 0.5 & 0.0 &  0.000000 &  0.000000\\
 0.1 & 1.0 & 0.0 &  0.000000 &  0.000000\\
 0.1 & 0.1 & 30.0 &  0.003124 &  0.003124\\
 0.1 & 0.5 & 30.0 & -0.000525 & -0.000525\\
 0.1 & 1.0 & 30.0 & -0.000706 & -0.000706
\end{tabular}

\vspace*{0.2cm}
Stokes parameter $V$ \\
\begin{tabular}{r r r r r}
$\mu_0$ & $\mu$ & $\phi-\phi_0$ & de Haan et al.\ & \pymiedap \\ \hline
0.5 & 0.1 & 0.0 &  0.000000 &  0.000000\\
 0.5 & 0.5 & 0.0 &  0.000000 &  0.000000\\
 0.5 & 1.0 & 0.0 &  0.000000 &  0.000000\\
 0.5 & 0.1 & 30.0 &  0.000038 &  0.000039\\
 0.5 & 0.5 & 30.0 &  0.000017 &  0.000018\\
 0.5 & 1.0 & 30.0 &  0.000000 &  0.000000\\
 0.1 & 0.1 & 0.0 &  0.000000 &  0.000000\\
 0.1 & 0.5 & 0.0 &  0.000000 &  0.000000\\
 0.1 & 1.0 & 0.0 &  0.000000 &  0.000000\\
 0.1 & 0.1 & 30.0 &  0.000012 &  0.000012\\
 0.1 & 0.5 & 30.0 &  0.000007 &  0.000007\\
 0.1 & 1.0 & 30.0 &  0.000000 &  0.000000
\end{tabular}
\caption{Stokes parameters $I$, $Q$, $U$, and $V$ of the locally
reflected light for model~1 (see Sect.~\ref{sect7.1}) 
(a 1--layer atmosphere with water-haze L-aerosol and a black surface), 
as listed in Table~5 of \citet{1987A&A...183..371D}, and as 
calculated using \pymiedap with 40 Gaussian abscissae.}
\label{benchmark-modA}
\end{table}

\begin{table}
Stokes parameter $I$ \\
\begin{tabular}{r r r r r}
$\mu_0$ & $\mu$ & $\phi-\phi_0$ & de Haan et al.\ & \pymiedap \\ \hline
0.5 & 0.1 & 0.0 &  0.532950 &  0.532930\\
 0.5 & 0.5 & 0.0 &  0.208430 &  0.208422\\
 0.5 & 1.0 & 0.0 &  0.093680 &  0.093680\\
 0.5 & 0.1 & 30.0 &  0.418140 &  0.418131\\
 0.5 & 0.5 & 30.0 &  0.184970 &  0.184974\\
 0.5 & 1.0 & 30.0 &  0.093680 &  0.093680\\
 0.1 & 0.1 & 0.0 &  0.522770 &  0.522887\\
 0.1 & 0.5 & 0.0 &  0.106590 &  0.106586\\
 0.1 & 1.0 & 0.0 &  0.026009 &  0.026009\\
 0.1 & 0.1 & 30.0 &  0.276300 &  0.276338\\
 0.1 & 0.5 & 30.0 &  0.083628 &  0.083626\\
 0.1 & 1.0 & 30.0 &  0.026009 &  0.026009
\end{tabular}

\vspace*{0.2cm}
Stokes parameter $Q$ \\
\begin{tabular}{r r r r r}
$\mu_0$ & $\mu$ & $\phi-\phi_0$ & de Haan et al.\ & \pymiedap \\ \hline
0.5 & 0.1 & 0.0 & -0.028340 & -0.028339\\
 0.5 & 0.5 & 0.0 & -0.036299 & -0.036298\\
 0.5 & 1.0 & 0.0 & -0.024156 & -0.024156\\
 0.5 & 0.1 & 30.0 & -0.000058 & -0.000057\\
 0.5 & 0.5 & 30.0 & -0.019649 & -0.019649\\
 0.5 & 1.0 & 30.0 & -0.012078 & -0.012078\\
 0.1 & 0.1 & 0.0 &  0.011506 &  0.011509\\
 0.1 & 0.5 & 0.0 & -0.005186 & -0.005185\\
 0.1 & 1.0 & 0.0 & -0.014984 & -0.014984\\
 0.1 & 0.1 & 30.0 &  0.034368 &  0.034376\\
 0.1 & 0.5 & 30.0 &  0.003839 &  0.003840\\
 0.1 & 1.0 & 30.0 & -0.007492 & -0.007492
\end{tabular}

\vspace*{0.2cm}
Stokes parameter $U$ \\
\begin{tabular}{r r r r r}
$\mu_0$ & $\mu$ & $\phi-\phi_0$ & de Haan et al.,\ & \pymiedap \\ \hline
0.5 & 0.1 & 0.0 &  0.000000 &  0.000000\\
 0.5 & 0.5 & 0.0 &  0.000000 &  0.000000\\
 0.5 & 1.0 & 0.0 &  0.000000 &  0.000000\\
 0.5 & 0.1 & 30.0 & -0.073105 & -0.073105\\
 0.5 & 0.5 & 30.0 & -0.041401 & -0.041401\\
 0.5 & 1.0 & 30.0 & -0.020920 & -0.020919\\
 0.1 & 0.1 & 0.0 &  0.000000 &  0.000000\\
 0.1 & 0.5 & 0.0 &  0.000000 &  0.000000\\
 0.1 & 1.0 & 0.0 &  0.000000 &  0.000000\\
 0.1 & 0.1 & 30.0 & -0.016042 & -0.016043\\
 0.1 & 0.5 & 30.0 & -0.014492 & -0.014492\\
 0.1 & 1.0 & 30.0 & -0.012976 & -0.012976
\end{tabular}

\vspace*{0.2cm}
Stokes parameter $V$ \\
\begin{tabular}{r r r r r}
$\mu_0$ & $\mu$ & $\phi-\phi_0$ & de Haan et al.\ & \pymiedap \\ \hline
0.5 & 0.1 & 0.0 &  0.000000 &  0.000000\\
 0.5 & 0.5 & 0.0 &  0.000000 &  0.000000\\
 0.5 & 1.0 & 0.0 &  0.000000 &  0.000000\\
 0.5 & 0.1 & 30.0 &  0.000106 &  0.000101\\
 0.5 & 0.5 & 30.0 &  0.000040 &  0.000036\\
 0.5 & 1.0 & 30.0 &  0.000000 &  0.000000\\
 0.1 & 0.1 & 0.0 &  0.000000 &  0.000000\\
 0.1 & 0.5 & 0.0 &  0.000000 &  0.000000\\
 0.1 & 1.0 & 0.0 &  0.000000 &  0.000000\\
 0.1 & 0.1 & 30.0 &  0.000027 &  0.000027\\
 0.1 & 0.5 & 30.0 &  0.000017 &  0.000017\\
 0.1 & 1.0 & 30.0 &  0.000000 &  0.000000
\end{tabular}
\caption{Similar to Table~1, except for model 2 (see 
Sect.~\ref{sect7.1}) (a 2--layer atmosphere with gas and 
water-haze L-aerosol, and a surface with albedo 0.1), 
as listed in Table~9 of \citet{1987A&A...183..371D} and as 
calculated using \pymiedap with 40 Gaussian abscissae.}
\label{benchmark-modB}
\end{table}


\begin{table}
\begin{tabular}{r | c c | c c}
\hline
\multirow{2}{*}{$\alpha$ \hspace*{0.05cm}} & 
\multicolumn{2}{c|}{$F$} &
\multicolumn{2}{c}{$P_{\rm s}$} \\
& Stam et al.\ & \pymiedapx & Stam et al. & \pymiedap \\ \hline
0.0 & 0.6471 & 0.6469 & 0.0000 & 0.0000\\
5.0 & 0.6424 & 0.6422 & 0.0021 & 0.0020\\
10.0 & 0.6299 & 0.6296 & 0.0081 & 0.0080\\
15.0 & 0.6108 & 0.6106 & 0.0179 & 0.0179\\
20.0 & 0.5861 & 0.5859 & 0.0315 & 0.0315\\
25.0 & 0.5570 & 0.5568 & 0.0487 & 0.0487\\
30.0 & 0.5245 & 0.5244 & 0.0693 & 0.0693\\
35.0 & 0.4898 & 0.4896 & 0.0930 & 0.0930\\
40.0 & 0.4536 & 0.4535 & 0.1195 & 0.1195\\
45.0 & 0.4171 & 0.4169 & 0.1483 & 0.1484\\
50.0 & 0.3808 & 0.3807 & 0.1789 & 0.1790\\
55.0 & 0.3455 & 0.3454 & 0.2105 & 0.2106\\
60.0 & 0.3118 & 0.3117 & 0.2422 & 0.2423\\
65.0 & 0.2799 & 0.2798 & 0.2730 & 0.2730\\
70.0 & 0.2501 & 0.2501 & 0.3015 & 0.3016\\
75.0 & 0.2226 & 0.2226 & 0.3266 & 0.3267\\
80.0 & 0.1975 & 0.1975 & 0.3469 & 0.3470\\
85.0 & 0.1745 & 0.1746 & 0.3613 & 0.3614\\
90.0 & 0.1537 & 0.1538 & 0.3689 & 0.3690\\
95.0 & 0.1349 & 0.1350 & 0.3690 & 0.3691\\
100.0 & 0.1179 & 0.1179 & 0.3615 & 0.3616\\
105.0 & 0.1024 & 0.1025 & 0.3466 & 0.3466\\
110.0 & 0.0883 & 0.0884 & 0.3249 & 0.3250\\
115.0 & 0.0755 & 0.0756 & 0.2976 & 0.2977\\
120.0 & 0.0638 & 0.0639 & 0.2659 & 0.2659\\
125.0 & 0.0531 & 0.0532 & 0.2311 & 0.2311\\
130.0 & 0.0434 & 0.0435 & 0.1946 & 0.1947\\
135.0 & 0.0347 & 0.0348 & 0.1579 & 0.1579\\
140.0 & 0.0269 & 0.0270 & 0.1220 & 0.1220\\
145.0 & 0.0201 & 0.0202 & 0.0882 & 0.0881\\
150.0 & 0.0143 & 0.0144 & 0.0573 & 0.0571\\
155.0 & 0.0095 & 0.0096 & 0.0302 & 0.0299\\
160.0 & 0.0058 & 0.0058 & 0.0079 & 0.0071\\
165.0 & 0.0031 & 0.0031 & -0.0088 & -0.0095\\
170.0 & 0.0012 & 0.0012 & -0.0184 & -0.0221\\
175.0 & 0.0003 & 0.0002 & -0.0186 & -0.0270\\
180.0 & 0.0000 & 0.0000 & 0.0000 & 0.0000
\end{tabular}
\caption{The flux and degree of polarization of light that is reflected by a 
         planet with a gaseous atmosphere with $b= 5.75$, bounded below by a 
         black surface, as a function of $\alpha$, as computed by \citet{Stam2006}
         and by using \pymiedap. The number of Gaussian abscissae, $N_{\rm G}$, 
         is 20 and $N_{\rm eq}=100$. The fluxes have been normalized such that 
         $F(0^\circ)= A_{\rm G}$.}
\label{gas-giant-1}
\end{table}


\begin{table}
\begin{tabular}{r | r c | r c}
\hline
\multirow{2}{*}{$\alpha$ \hspace*{0.05cm}} & 
\multicolumn{2}{c|}{$F$} &
\multicolumn{2}{c}{$P_{\rm s}$} \\
& Stam et al.\ & \pymiedapx & Stam et al. & \pymiedap \\ \hline
0.0 & 0.6688 & 0.6676 & 0.0000 & 0.0000\\
5.0 & 0.6512 & 0.6542 & 0.0023 & -0.0000\\
10.0 & 0.6439 & 0.6361 & -0.0047 & -0.0003\\
15.0 & 0.6128 & 0.6113 & 0.0083 & 0.0095\\
20.0 & 0.5866 & 0.5857 & 0.0272 & 0.0229\\
25.0 & 0.5593 & 0.5580 & 0.0496 & 0.0462\\
30.0 & 0.5294 & 0.5288 & 0.0691 & 0.0710\\
35.0 & 0.4956 & 0.4957 & 0.0860 & 0.0901\\
40.0 & 0.4590 & 0.4589 & 0.1040 & 0.1077\\
45.0 & 0.4215 & 0.4212 & 0.1253 & 0.1285\\
50.0 & 0.3849 & 0.3844 & 0.1494 & 0.1527\\
55.0 & 0.3497 & 0.3492 & 0.1749 & 0.1784\\
60.0 & 0.3163 & 0.3157 & 0.2005 & 0.2043\\
65.0 & 0.2847 & 0.2841 & 0.2251 & 0.2292\\
70.0 & 0.2550 & 0.2545 & 0.2475 & 0.2519\\
75.0 & 0.2275 & 0.2269 & 0.2669 & 0.2716\\
80.0 & 0.2020 & 0.2015 & 0.2824 & 0.2874\\
85.0 & 0.1787 & 0.1782 & 0.2932 & 0.2985\\
90.0 & 0.1575 & 0.1571 & 0.2986 & 0.3042\\
95.0 & 0.1382 & 0.1378 & 0.2980 & 0.3037\\
100.0 & 0.1208 & 0.1204 & 0.2912 & 0.2968\\
105.0 & 0.1050 & 0.1047 & 0.2782 & 0.2836\\
110.0 & 0.0907 & 0.0905 & 0.2593 & 0.2643\\
115.0 & 0.0778 & 0.0776 & 0.2353 & 0.2397\\
120.0 & 0.0662 & 0.0661 & 0.2073 & 0.2111\\
125.0 & 0.0557 & 0.0557 & 0.1765 & 0.1794\\
130.0 & 0.0463 & 0.0464 & 0.1444 & 0.1461\\
135.0 & 0.0380 & 0.0381 & 0.1126 & 0.1132\\
140.0 & 0.0306 & 0.0309 & 0.0827 & 0.0816\\
145.0 & 0.0242 & 0.0246 & 0.0561 & 0.0530\\
150.0 & 0.0187 & 0.0192 & 0.0340 & 0.0293\\
155.0 & 0.0140 & 0.0144 & 0.0171 & 0.0125\\
160.0 & 0.0101 & 0.0103 & 0.0061 & 0.0031\\
165.0 & 0.0068 & 0.0069 & 0.0017 & 0.0006\\
170.0 & 0.0045 & 0.0039 & -0.0037 & 0.0019\\
175.0 & 0.0054 & 0.0032 & -0.0042 & -0.0043\\
180.0 & 0.0000 & 0.0000 & 0.0000 & 0.0000
\end{tabular}
\caption{Similar to Table~\ref{gas-giant-1}, except for a model atmosphere 
         with model~D aerosol \citep[][]{1984A&A...131..237D} added, such that
         $b=9.0$. For these computations, $N_{\rm G}=50$.}
\label{gas-giant-2}
\end{table}

\section{Summary}
\label{sect-summary}

We presented \pymiedapx, a modular Python--based tool to compute the total and 
polarized fluxes of light that is reflected by (exo)planets (or moons) 
with locally 
horizontally homogeneous, plane--parallel atmospheres bounded below by 
a horizontally homogeneous, flat surface. The atmospheres can be vertically
inhomogeneous. Horizontally inhomogeneous planets are modelled by assigning
different atmosphere-surface combinations to different regions on the planet.
The Fortran radiative transfer algorithm is based on the adding--doubling method 
as described by \citet{1987A&A...183..371D}, and fully includes linear 
and circular polarization for all orders of scattering.
The single scattering of light by atmospheric aerosols is computed using
Mie--scattering, based on \citet{1984A&A...131..237D}.

\pymiedap has a two-step approach: first, the adding--doubling radiative transfer
computations provide files with Fourier coefficients of the expansion of the 
local reflection matrix of the model planetary atmosphere and surface, 
and, second, the Fourier coefficients are used to efficiently compute the
locally reflected Stokes vectors for every given geometry. The latter 
Stokes parameters can be summed up to provide the disk--integrated Stokes
parameters of the reflected starlight. By storing 
the Fourier--coefficient files for later use, significant amounts of 
computing time can be saved in the computation of the reflected light vectors.

The modular aspect of \pymiedap allows users to define an atmosphere-surface
model and to compute spatially resolved and/or disk--integrated signals 
of a planet at a range of phase angle in a single function call.
\pymiedap can straightforwardly be used to model signals of horizontally 
inhomogeneous planets by assigning different atmosphere--surface models to 
different regions on a planet. Four pre-defined cloud types or 'masks'
are included in the code.
The modular aspect of the code also allows for
step-by-step computations, for users who wish to perform more complicated
cases. 

\pymiedap is distributed under the GNU GPL license and we invite
interested users to suggest improvements or extensions to broaden the 
application of the code.


\paragraph{Acknowledgements} L.R.\ acknowledges the support of the Dutch
Scientific Organization (NWO) through the PEPSci network of planetary and
exoplanetary science. The authors thank Gourav Mahapatra and Ashwyn Groot, who were kind enough to 
test the code before its public release.


\bibliographystyle{aa}
\bibliography{refs_dmstam}

\begin{thebibliography}{42}
\expandafter\ifx\csname natexlab\endcsname\relax\def\natexlab#1{#1}\fi

\bibitem[{{Aben} {et~al.}(1997){Aben}, {Helderman}, {Stam}, \&
  {Stammes}}]{1997ABEN}
{Aben}, I., {Helderman}, F., {Stam}, D., \& {Stammes}, P. 1997, in
  Polarization: Measurement, Analysis, and Remote Sensing. Proceedings SPIE
  {\bf 3121}, ed. D.~{Goldstein} \& R.~{Chipman}, 446--451

\bibitem[{{Bates}(1984)}]{1984P&SS...32..785B}
{Bates}, D.~R. 1984, \planss, 32, 785

\bibitem[{Bideau-Mehu {et~al.}(1973)Bideau-Mehu, Guern, Abjean, \&
  Johannin-Gilles}]{Bideau-Mehu1973}
Bideau-Mehu, A., Guern, Y., Abjean, R., \& Johannin-Gilles, A. 1973, Optics
  Communications, 9, 432

\bibitem[{Boesche {et~al.}(2009)Boesche, Stammes, \& Bennartz}]{Boesche2009}
Boesche, E., Stammes, P., \& Bennartz, R. 2009, Journal of Quantitative
  Spectroscopy and Radiative Transfer, 110, 223

\bibitem[{{Bohren} \& {Huffman}(1983)}]{Bohren1983}
{Bohren}, C. \& {Huffman}, D. 1983, {Absorption and scattering of light by
  small particles} (Wiley), 477--482

\bibitem[{Ciddor(1996)}]{Ciddor1996}
Ciddor, P.~E. 1996, Applied optics, 35, 1566

\bibitem[{Cotton {et~al.}(2017)Cotton, Marshall, Bailey, Kedziora-Chudczer,
  Bott, Marsden, \& Carter}]{Cotton2017}
Cotton, D.~V., Marshall, J.~P., Bailey, J., {et~al.} 2017, Monthly Notices of
  the Royal Astronomical Society, 467, 873

\bibitem[{{de Haan} {et~al.}(1987){de Haan}, {Bosma}, \&
  {Hovenier}}]{1987A&A...183..371D}
{de Haan}, J.~F., {Bosma}, P.~B., \& {Hovenier}, J.~W. 1987, \aap, 183, 371

\bibitem[{{de Rooij} \& {van der Stap}(1984)}]{1984A&A...131..237D}
{de Rooij}, W.~A. \& {van der Stap}, C.~C.~A.~H. 1984, \aap, 131, 237

\bibitem[{{Deirmendjian}(1969)}]{1969esos.book.....D}
{Deirmendjian}, D. 1969, {Electromagnetic scattering on spherical
  polydispersions.}

\bibitem[{Fauchez {et~al.}(2017)Fauchez, Rossi, \& Stam}]{Fauchez2017}
Fauchez, T., Rossi, L., \& Stam, D.~M. 2017, The Astrophysical Journal, 842, 41

\bibitem[{{Grainger} \& {Ring}(1962)}]{1962Natur.193..762W}
{Grainger}, J.~F. \& {Ring}, J. 1962, \nat, 193, 762

\bibitem[{{Hansen} \& {Hovenier}(1974)}]{Hansen1974}
{Hansen}, J.~E. \& {Hovenier}, J.~W. 1974, Journal of Atmospheric Sciences, 31,
  1137

\bibitem[{{Hansen} \& {Travis}(1974)}]{1974SSRv...16..527H}
{Hansen}, J.~E. \& {Travis}, L.~D. 1974, Space Science Reviews, 16, 527

\bibitem[{{Hovenier} \& {Stam}(2005)}]{2005HovenierStam}
{Hovenier}, J.~W. \& {Stam}, D.~M. 2005, Journal of Quantitative Spectroscopy
  and Radiative Transfer (in press)

\bibitem[{{Hovenier} {et~al.}(2004){Hovenier}, {van der Mee}, \&
  {Domke}}]{2004Hovenier}
{Hovenier}, J.~W., {van der Mee}, C., \& {Domke}, H. 2004, {Transfer of
  Polarized Light in Planetary Atmospheres; Basic Concepts and Practical
  Methods} (Kluwer, Dordrecht; Springer, Berlin)

\bibitem[{{Hovenier} \& {van der Mee}(1983)}]{1983A&A...128....1H}
{Hovenier}, J.~W. \& {van der Mee}, C.~V.~M. 1983, \aap, 128, 1

\bibitem[{{Kemp} {et~al.}(1987){Kemp}, {Henson}, {Steiner}, \&
  {Powell}}]{1987Natur.326..270K}
{Kemp}, J.~C., {Henson}, G.~D., {Steiner}, C.~T., \& {Powell}, E.~R. 1987,
  \nat, 326, 270

\bibitem[{Mansfield \& Peck(1969)}]{Mansfield1969}
Mansfield, C.~R. \& Peck, E.~R. 1969, JOSA, 59, 199

\bibitem[{{Mishchenko} {et~al.}(1994){Mishchenko}, {Lacis}, \&
  {Travis}}]{1994JQSRT..51..491M}
{Mishchenko}, M.~I., {Lacis}, A.~A., \& {Travis}, L.~D. 1994, Journal of
  Quantitative Spectroscopy and Radiative Transfer, 51, 491

\bibitem[{{Mishchenko} {et~al.}(2002){Mishchenko}, {Travis}, \&
  {Lacis}}]{2002sael.book.....M}
{Mishchenko}, M.~I., {Travis}, L.~D., \& {Lacis}, A.~A. 2002, {Scattering,
  absorption, and emission of light by small particles}

\bibitem[{{Mu{\~n}oz} {et~al.}(2012){Mu{\~n}oz}, {Moreno}, {Guirado},
  {Dabrowska}, {Volten}, \& {Hovenier}}]{2012JQSRT.113..565M}
{Mu{\~n}oz}, O., {Moreno}, F., {Guirado}, D., {et~al.} 2012, \jqsrt, 113, 565

\bibitem[{Peck \& Huang(1977)}]{Peck1977}
Peck, E.~R. \& Huang, S. 1977, JOSA, 67, 1550

\bibitem[{Peck \& Khanna(1966)}]{Peck1966}
Peck, E.~R. \& Khanna, B.~N. 1966, JOSA, 56, 1059

\bibitem[{Peterson(2009)}]{Peterson2009}
Peterson, P. 2009, International Journal of Computational Science and
  Engineering, 4, 296

\bibitem[{Pierrehumbert(2011)}]{Pierrehumbert2011}
Pierrehumbert, R.~T. 2011, The Astrophysical Journal Letters, 726, L8

\bibitem[{{Press} {et~al.}(1992){Press}, {Teukolsky}, {Vetterling}, \&
  {Flannery}}]{1992nrfa.book.....P}
{Press}, W.~H., {Teukolsky}, S.~A., {Vetterling}, W.~T., \& {Flannery}, B.~P.
  1992, {Numerical recipes in FORTRAN. The art of scientific computing}

\bibitem[{Rossi \& Stam(2017)}]{Rossi2017}
Rossi, L. \& Stam, D. 2017, Astronomy \& Astrophysics

\bibitem[{{Seager} {et~al.}(2000){Seager}, {Whitney}, \&
  {Sasselov}}]{2000ApJ...540..504S}
{Seager}, S., {Whitney}, B.~A., \& {Sasselov}, D.~D. 2000, \apj, 540, 504

\bibitem[{{Sneep} \& {Ubachs}(2005)}]{2005JQSRT..92..293S}
{Sneep}, M. \& {Ubachs}, W. 2005, \jqsrt, 92, 293

\bibitem[{{Stam}(2008)}]{2008A&A...482..989S}
{Stam}, D.~M. 2008, \aap, 482, 989

\bibitem[{{Stam} {et~al.}(2002){Stam}, {Aben}, \&
  {Helderman}}]{2002JGRD.107t.AAC1S}
{Stam}, D.~M., {Aben}, I., \& {Helderman}, F. 2002, Journal of Geophysical
  Research (Atmospheres), AAC 1

\bibitem[{{Stam} {et~al.}(2000){Stam}, {De Haan}, {Hovenier}, \&
  {Aben}}]{2000JGR...10522379S}
{Stam}, D.~M., {De Haan}, J.~F., {Hovenier}, J.~W., \& {Aben}, I. 2000, \jgr,
  22379

\bibitem[{{Stam} {et~al.}(2006){Stam}, {de Rooij}, {Cornet}, \&
  {Hovenier}}]{Stam2006}
{Stam}, D.~M., {de Rooij}, W.~A., {Cornet}, G., \& {Hovenier}, J.~W. 2006,
  \aap, 452, 669

\bibitem[{{Stam} \& {Hovenier}(2005)}]{2005A&A...444..275S}
{Stam}, D.~M. \& {Hovenier}, J.~W. 2005, \aap, 444, 275

\bibitem[{{Stam} {et~al.}(2004){Stam}, {Hovenier}, \&
  {Waters}}]{2004A&A...428..663S}
{Stam}, D.~M., {Hovenier}, J.~W., \& {Waters}, L.~B.~F.~M. 2004, \aap, 428, 663

\bibitem[{{Stammes} {et~al.}(1994){Stammes}, {Kuik}, \& {de
  Haan}}]{1994STAMMES}
{Stammes}, P., {Kuik}, F., \& {de Haan}, J. 1994, in Proceedings PIERS 1994,
  Kluwer Acad., Dordrecht, ed. B.~e.~a. {Arbesser-Rastburg}, 2255--2259

\bibitem[{{Turbet} {et~al.}(2016){Turbet}, {Leconte}, {Selsis}, {Bolmont},
  {Forget}, {Ribas}, {Raymond}, \& {Anglada-Escudé}}]{Turbet2016}
{Turbet}, M., {Leconte}, J., {Selsis}, F., {et~al.} 2016, A\&A, 596, A112

\bibitem[{{van de Hulst}(1980)}]{1980vandeHulst}
{van de Hulst}, H.~C. 1980, {Multiple Light Scattering, Tables, Formulas, and
  Applications, {\rm Vols. 1 and 2}} (Academic Press, New York.)

\bibitem[{Yang {et~al.}(2013)Yang, Cowan, \& Abbot}]{Yang2013}
Yang, J., Cowan, N.~B., \& Abbot, D.~S. 2013, The Astrophysical Journal
  Letters, 771, L45

\bibitem[{{Young}(1981)}]{1981ApOpt..20..533Y}
{Young}, A.~T. 1981, \ao, 20, 533

\bibitem[{{Yurkin} \& {Hoekstra}(2011)}]{2011JQSRT.112.2234Y}
{Yurkin}, M.~A. \& {Hoekstra}, A.~G. 2011, \jqsrt, 112, 2234

\end{thebibliography}

\appendix


\section{Fourier file formats}
\label{fourier-files}

A Fourier--coefficients file for a given model atmosphere-surface combination
has the following format: \\

\noindent The first lines contain comments, including a reference, and they have
details on the model atmosphere and surface. The number of these lines depends
on the number of layers in the model atmosphere, 
but they are all preceded by a '\#'. In the following, we will assume the
number of comment lines is $N$. \\

\noindent Line $N + 1$ contains a number to describe the size of the reflected 
light vectors: '1' indicates only $I$, '3' indicates $I$, $Q$, and $U$,
and '4' indicates $I$, $Q$, $U$, and $V$. \\

\noindent Line $N + 2$ contains the number of Gaussian abscissae $G$ plus
the supplemented value 1.0. It thus contains the value $G+1$. \\

\noindent Lines $N + 3$ up to and including $N + 4 + N_{\rm G}$ 
contain the $N_{\rm G}$ Gaussian abscissae
(the cosines of the corresponding illumination and viewing zenith angles) 
plus the supplemented value 1.0, and the corresponding Gaussian weights.
For the supplemented value, this weight is set equal to 1.0. \\

\noindent Starting with line $N + N_{\rm G} + 5$, the elements of the 
first column of the local 
reflection matrix ${\bf R}$, i.e. $R^m_{11}$, $R^m_{21}$, $R^m_{31}$, and $R^m_{41}$
(see Eqs.~\ref{eq1}-\ref{eq4}) are listed, 
with $m$ the number of the Fourier term
($0 \leq m \leq M$, with $M$ the number of the last Fourier term).
Elements $R^m_{21}$ and $R^m_{31}$ are only listed if the
polarized fluxes have actually been calculated, and element $R^m_{41}$
is only listed if the circularly polarized flux has also been calculated.
The matrix elements depend not only on the number of the Fourier term, $m$, but
also on the illumination and viewing zenith angles, i.e. on $\mu$ and $\mu_0$.
With $N_{\rm G}$ 'true' Gaussian abscissae and 1 supplemented value, 
we have $(N_{\rm G}+1)^2$ combinations of $\mu$ and $\mu_0$.

Each line has the format: $m$, $i$, $j$, $R^m_{k1}$, with $k$ equal to 1, 3, or 4,
with $i$ the number of the Gaussian abscissae representing 
$\mu$ ($1 \leq i \leq N_{\rm G}+1$), and 
with $j$ the number of the Gaussian abscissae representing $\mu_0$ 
($1 \leq j \leq N_{\rm G}+1$).
Each file thus has $(M+1)(N_{\rm G}+1)^2$ lines with 1 to 4 elements of the 
first column of 
the local reflection matrix ${\bf R}$. For a purely gaseous atmosphere, 
$M=2$, and given
$N_{\rm G}=20$, the total number of lines with matrix elements is thus 1,323.
Model atmospheres with aerosol and/or cloud particles will usually require
much larger values for $M$, and will thus yield much larger data files.


\section{Computation of the local angles}
\label{appendix-angles}

We describe here equations that are used to compute the local illumination and viewing 
angles $\theta_{0i}$, $\theta_i$, $\beta_i$ and $\phi_i - \phi_{0i}$ for a 
pixel $i$ in the pixel grid. First, we define the planetocentric reference 
frame $(\mathbf{u_x},\mathbf{u_y},\mathbf{u_z})$, where $\mathbf{u_x}$ and 
$\mathbf{u_y}$ lie in the plane of the observer's sky, while $\mathbf{u_z}$ 
points towards the observer.

Assume that the coordinates of pixel $i$ in the plane of the sky 
are given by $(x_i,y_i)$. Assuming that the planet is spherical with a radius 
equal to one, we know that
the 3D coordinates of the projected pixel centre on the planet are 
$(x_i,y_i,z_i)$ with $z_i = (x_i^2 + y_i^2)^{1/2}$.

The local zenith direction for pixel $i$ is given by vector 
\begin{equation}
   \mathbf{r_{cp}}= 
   \left[ 
   \begin{array}{c}
          x_i \\ y_i \\ z_i 
   \end{array} 
   \right],
\end{equation}
and the vector pointing to the star is given by
\begin{equation}
   \mathbf{r_{cs}}= 
   \left[ 
   \begin{array}{c}
          \sin{\alpha} \\ 0 \\ \cos{\alpha} 
   \end{array} 
   \right],
\end{equation}
where $\alpha$ is the planetary phase angle, i.e.\ the angle between the
direction to the observer and the direction to the star as measured from the 
centre of the planet. 
The local solar/stellar zenith angle is thus given by
\begin{equation}
   \theta_{0i} = \mathbf{r_{cp}} \cdot \mathbf{r_{cs}} 
               = \arccos(x_i \sin{\alpha} + z_i \cos{\alpha}),
\end{equation}
and, since unit vector $\mathbf{u_z}$ is pointing towards the observer, 
the local viewing zenith angle is given by
\begin{equation}
   \theta_i = \arccos{z_i}.
\end{equation}
The local azimuthal difference angle $\phi_i - \phi_{0i}$ can be computed
using the spherical law of cosines:
\begin{equation}
\cos{\alpha} = \cos{\theta_{0i}} \cos{\theta_i} + \sin{\theta_{0i}} \sin{\theta_i} 
               \cos(\phi_i - \phi_{0i}),
\end{equation}
and thus
\begin{equation}
   \phi_i - \phi_{0i} = \arccos \left( \frac{\cos{\alpha} - 
                        \cos{\theta_{0i}} \cos{\theta_i}}{\sin{\theta_{0i}} 
                        \sin{\theta_i}} \right).
\end{equation}

For $0 < \alpha < \pi$, the local rotation angle $\beta_i$ that is used to 
rotate computed Stokes parameters defined with respect to the local meridian 
planet to the planetary scattering plane is given by
\begin{equation}
  \beta_i=
  \begin{cases}
	\arctan{ \left( y_i/x_i \right)} & 
           \mathrm{if} \hspace*{0.2cm} x_i \hspace*{0.05cm} y_i \geq 0 \\
	180^\circ + \arctan{\left( y_i/x_i \right)} & 
           \mathrm{if} \hspace*{0.2cm} x_i \hspace*{0.05cm} y_i < 0 \\
  \end{cases}
\end{equation}
For $\pi < \alpha < 2\pi$, rotation angle $\beta_i$ is given by 
\begin{equation}
  \beta_i =
  \begin{cases}
	180^\circ - \arctan{ \left( y_i/x_i \right)} & 
    \mathrm{if} \hspace*{0.2cm} x_i \hspace*{0.05cm} y_i \geq 0\\
	- \arctan{\left( y_i/x_i \right)} & 
    \mathrm{if} \hspace*{0.2cm} x_i \hspace*{0.05cm} y_i < 0\\
  \end{cases}
\end{equation}

\end{document}